\documentclass[epj]{svjour}
\usepackage{amssymb,graphics,graphicx,times,amsmath,color,hyperref}
\def\be{\begin{equation}}
\def\ee{\end{equation}}
\def\ba{\begin{array}}
\def\ea{\end{array}}
\def\bea{\begin{eqnarray}}
\def\eea{\end{eqnarray}}
\def\bi{\begin{itemize}}
\def\ei{\end{itemize}}

\def\half{{\textstyle{1\over2}}}

\begin{document}
\title{Constraints on the symmetry energy from observational probes of the neutron star crust}
\author{William G. Newton\inst{1} \and Joshua Hooker\inst{1} \and Michael Gearheart\inst{1} \and Kyleah Murphy \inst{1,2} \and De-Hua Wen\inst{1,3} \and Farrukh J. Fattoyev\inst{1} \and Bao-An Li\inst{1}}

%\and Farrukh Fattoyev\inst{1}

\offprints{}          % Insert a name or remove this line
\institute{$^1$Department of Physics and Astronomy, Texas A\&M University-Commerce, Commerce, TX 75249, USA \\
$^2$Umpqua Community College, Roseburg, Oregon, 97470, USA \\
$^3$Department of Physics, South China University of Technology, Guangzhou 510641, P. R. China}
\date{Received: date / Revised version: date}
\abstract{
A number of observed phenomena associated with individual neutron star systems or neutron star populations find explanations in models in which the neutron star crust plays an important role. We review recent work examining the sensitivity to the slope of the symmetry energy $L$ of such models, and constraints extracted on $L$ from confronting them with observations. We focus on six sets of observations and proposed explanations: (i) The cooling rate of the neutron star in Cassiopeia A, confronting cooling models which include enhanced cooling in the nuclear pasta regions of the inner crust, (ii) the upper limit of the observed periods of young  X-ray pulsars, confronting models of magnetic field decay in the crust caused by the high resistivity of the nuclear pasta layer, (iii) glitches from the Vela pulsar, confronting the paradigm that they arise due to a sudden re-coupling of the crustal neutron superfluid to the crustal lattice after a period during which they were decoupled due to vortex pinning, (iv) The frequencies of quasi-periodic oscillations in the X-ray tail of light curves from giant flares from soft gamma-ray repeaters, confronting models of torsional crust oscillations, (v) the upper limit on the frequency to which millisecond pulsars can be spun-up due to accretion from a binary companion, confronting models of the r-mode instability arising above a threshold frequency determined in part by the viscous dissipation timescale at the crust-core boundary, and (vi) the observations of precursor electromagnetic flares a few seconds before short gamma-ray bursts, confronting a model of crust shattering caused by resonant excitation of a crustal oscillation mode by the tidal gravitational field of a companion neutron star just before merger.
 % end of PACS codes
} %end of abstract
\authorrunning{W. G. Newton {\sl et al.}}
\maketitle
\section{Introduction}
\label{intro}

The problem of constraining the isospin-dependence of nuclear forces is a vigorous field bringing together nuclear physicists and astrophysicists in theoretical, experimental and observational studies. The motivations involve obtaining a better understanding of the structural and dynamical properties of neutron rich nuclei and neutron stars, and an improved microscopic understanding of the nucleon-nucleon interaction and its emergence as a residual strong interaction of QCD.

The interior structure of a neutron star (NS) is poorly constrained because of uncertainties in the high-density behavior of the nuclear equation-of-state (EOS) and the plethora of exotic phases that have been theoretically postulated to exist towards the center of the star. Direct observational probes of the core of the neutron star are limited to neutrino and gravitational wave emission, both of which are yet to be directly detected. The outer regions of the star are more readily accessible via electromagnetic radiation from the surface and magnetosphere, and provide probes of the physics of the neutron star crust and the EOS around nuclear saturation density. Such constraints are intrinsically interesting and help scaffold our understanding of the deeper layers of the star. In this review, we focus on a number of recently explored neutron star observables that probe crustal properties, focussing on their potential to provide constraints for the nuclear symmetry energy, the most uncertain part of the nuclear EOS at densities close to nuclear saturation.

%====================================== FIGURE 1 ===============================

\begin{figure*}
\resizebox{1.0\textwidth}{!}{
  \includegraphics{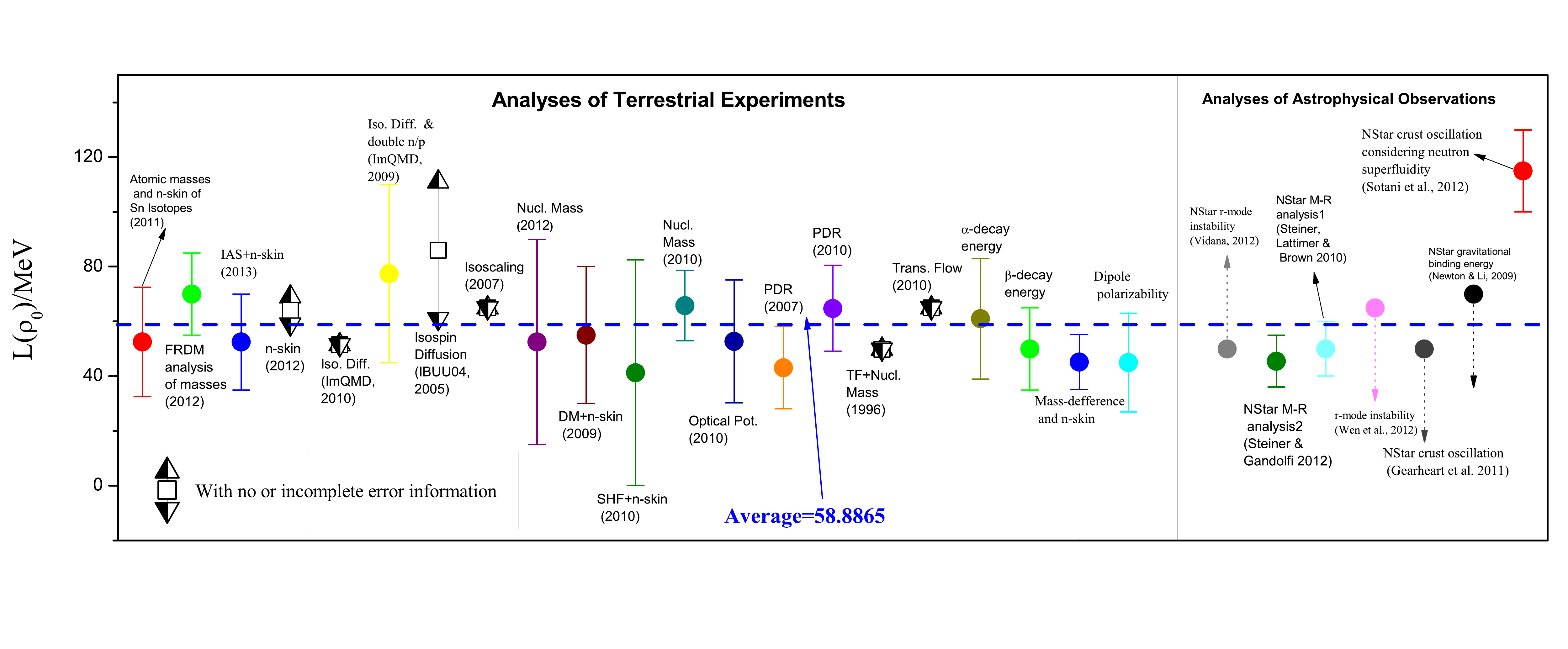}}
\caption{(Color online) Recent constraints on the slope of the symmetry energy $L$ from analysis of terrestrial experiments and astrophysical observations, some of which can be found summarized in this issue. Taken from \cite{Han2013}. The community average $L \approx$ 60 MeV should be taken only as guide to the favored values of $L$ that emerge from the wide variety of experimental evidence.}
\label{fig:1}       % Give a unique label
\end{figure*}

%==============================================================================

The isospin dependence of nuclear forces is manifest in dense nuclear matter as an energy cost of decreasing the proton fraction away from symmetric matter $N=Z$, a cost that is encoded in the symmetry energy as a function of density. There are many different notations for the symmetry energy and its expansion parameters around saturation density; let us introduce the one we use now.

The degree of isospin asymmetry in a system may be expressed locally in terms of the proton $n_{\rm p}$ and neutron $n_{\rm n}$ number densities (which might vary in space) via the local asymmetry parameter $\delta = (n_{\rm n} - n_{\rm p})/n$ where $n = n_{\rm n} + n_{\rm p}$ is the total baryon number density, or globally via the parameter $I = (N-Z)/A$, where $N,Z$ and $A$ are the total number of neutrons, protons and nucleons in the system. For uniform nuclear matter, both parameters $\delta$ and $I$ are identical. Nuclear matter with equal numbers of neutrons and protons ($\delta=0$) is referred to as symmetric nuclear matter (SNM); nuclear matter with $\delta=1$ is naturally referred to as pure neutron matter (PNM).  Nuclei on Earth contain closely symmetric nuclear matter at densities close to nuclear saturation density $\rho_0 \approx 2.6 \times 10^{14}$ g cm$^{-3} \equiv 0.16$ fm$^{-3} = n_0$. Nuclear experiments tend to constrain the behavior of the binding energy of symmetric nuclear matter, $E(n\sim n_0,\delta \sim0)$ to within relatively tight ranges, but direct \emph{ab initio} calculations there are extremely difficult. In contrast, experimental data directly probing $E(n\sim n_0,\delta \sim1)$ is impossible, but state-of-the-art $\emph{ab initio}$ calculations of PNM have led to significant constraints on $E(n < n_0,\delta \sim1)$. By expanding $E(n, \delta)$ about $\delta = 0$,
\be\label{eq:eos1}
	E(n,\delta) = E_{\rm 0}(n) + S(n)\delta^2 + ...,
\ee
\noindent we can define a useful quantity called the \emph{symmetry energy} 
\be\label{eq:symm}
S(n) = {1 \over 2}{\partial^2 E(n,\delta) \over \partial \delta^2}\bigg|_{\delta=0},
\ee
\noindent which encodes the change in the energy per particle of nuclear matter as one moves away from isospin symmetry. This allows extrapolation to the highly isospin asymmetric conditions in neutron stars. The simplest such extrapolation, referred to as the \emph{parabolic approximation} (PA) gives the relation
\be \label{PAapp}
E_{\rm PNM}(n) \approx E_{\rm 0}(n) + S(n).
\ee
Since our experimental constraints are dominated by results from densities close to $n_0$, it is customary to expand the symmetry energy about 
$\chi=0$ where $\chi = \frac{n-n_{\rm 0}}{3n_{\rm 0}}$, thus obtaining
\be\label{eq:eos3}
	S(n) = J + L \chi + \half K_{\rm sym} \chi^{2} + ..., %\frac{J_{sym}}{6}\chi^{3} + \frac{I_{sym}}{24}\chi^4 + ...
\ee

\noindent where $J$, $L$ and $K_{\rm sym}$ are the symmetry energy, its slope and its curvature at saturation density. The true values of the higher order symmetry energy parameters $L$, $K_{\rm sym}, ...$ are still somewhat uncertain, and the measurement of $L$ in particular has been the subject of an intense experimental campaign by the nuclear physics community in recent years using nuclear probes such as masses, neutron skins, nuclear electric dipole polarizability, collective motion and the dynamics of heavy ion collisions (see, e.g.\cite{Li2008,BTsang2012,Lattimer2013} for recent summaries). \emph{Ab initio} calculations of PNM with well defined theoretical errors offer additional constraints on $J$ and $L$ \cite{Schwenk2005,Hebeler2010,Gandolfi2009,Gandolfi2010,Gezerlis2013}. Both theory and experiment are generally in broad agreement that $L$ falls in the rather loose range $30 \lesssim L \lesssim 80$ MeV, although higher values in particular are not completely ruled out \cite{Fattoyev2013}. Fig.~1 shows a selection of experimental constraints on $L$, together with constraints inferred from astrophysical observation, some of which will be discussed in this review \cite{Han2013}.

Since neutron star matter contains a low fraction of protons, many inner crust and global stellar properties are sensitive to the symmetry energy parameters. To give a classic example, the pressure of PNM at saturation density is given in the parabolic approximation by $P_{\rm PNM}(n_0)$=$n_0L/3$. The pressure at the crust-core boundary and in the outer core is dominated by neutron pressure so a strong correlation should exist between the pressure in neutron stars near saturation density and $L$. Neutron star EOSs which have higher pressures at a particular density are often referred to as `stiff'; lower pressure EOSs are referred to as `soft'. Thus `stiff' EOSs at saturation density are associated with high values of $L$ and `soft' EOSs with low values of $L$. This fact leads to a strong correlation between the radii of neutron stars and the slope of the symmetry energy near saturation density \cite{Lattimer2001}.

%====================================== FIGURE 2 ===============================

\begin{figure}
\resizebox{0.5\textwidth}{!}{
  \includegraphics{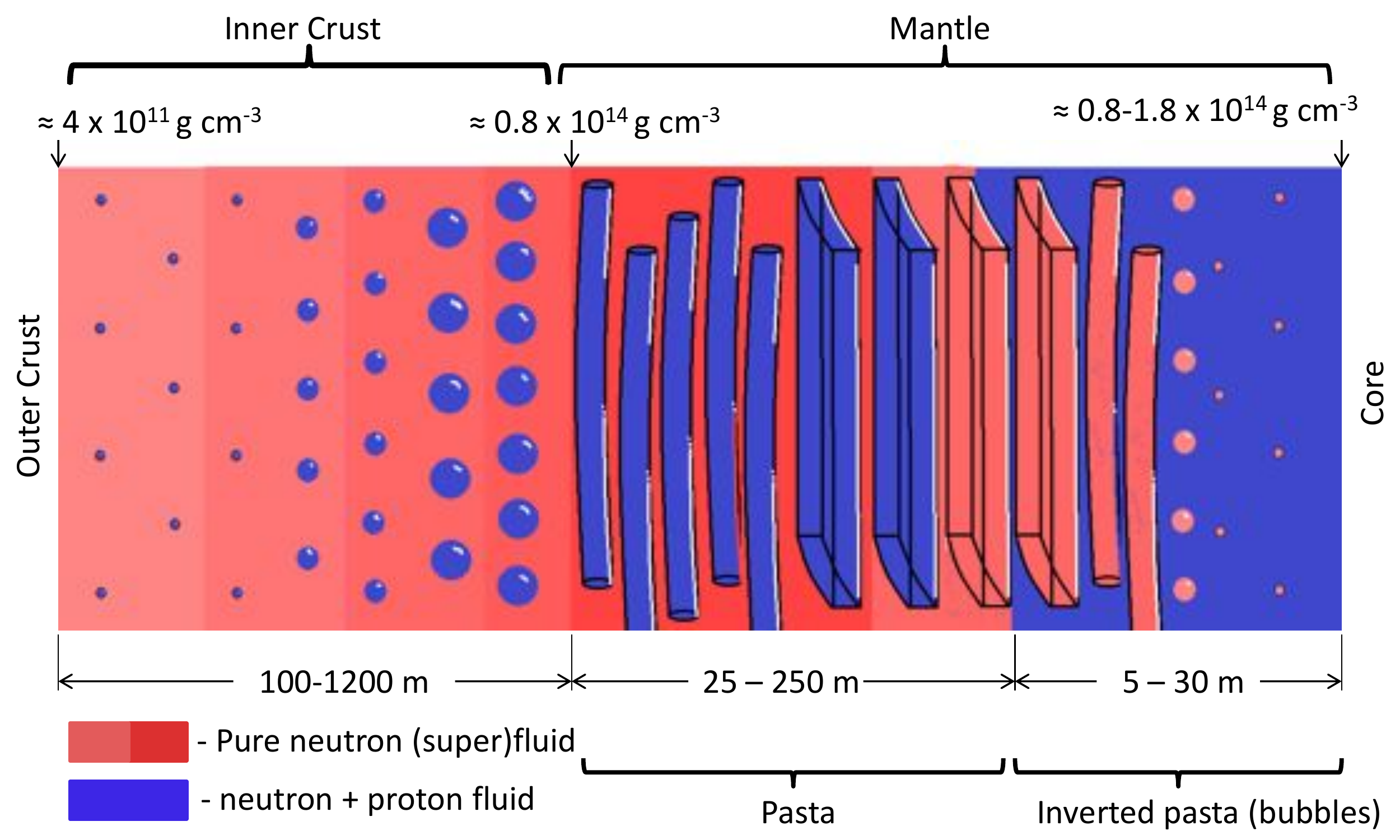}}
\caption{(Color online) Schematic depiction of the structure of the neutron star inner crust from the outer crust-inner crust boundary (left-hand side) to the inner crust-core boundary (right hand side). Taken from \cite{NewtonBook2012}.}
\label{fig:2}       % Give a unique label
\end{figure}

%==============================================================================

The crust of a neutron star is divided into two layers; the outer crust and inner crust. The microscopic structure of the crust is a crustal lattice of nuclei. The nuclei become increasingly more massive and neutron-rich with depth, immersed in a gas of relativistic electrons. The distinction between the two layers comes from the existence of an additional component of interstitial fluid neutrons in the inner crust. The absence of free neutrons in the outer crust allow a more direct connection between theoretical models and measurable exotic nuclear masses, and the composition and properties of the outer crust are substantially less uncertain than those of the inner crust (although uncertainty does still remain, and is connected in part with uncertainty in the nuclear symmetry energy \cite{RocaMaza2008}). A schematic picture of the structure of the inner crust is shown in Fig.~2. The free neutrons in the inner crust are expected to be in a superfluid state, which finds some observational support from studying the cooling of neutron stars in X-ray transients, after accretion induced heating of the crust \cite{Shternin2007,Brown2009}. At the base of the inner crust, competition between the inter-nuclear Coulomb energy and the surface energy of nuclei make it energetically favorable for the nuclei to form cylindrical, slab or cylindrical/spherical bubble shapes, a set of phases which go under the name ``nuclear pasta'' \cite{Ravenhall1983,Hashimoto1984}. Analogous changes in the microscopic structure of certain terrestrial soft condensed matter systems such as surfactants are known, and we might expect the nuclear pasta phases to exhibit similarly rich and complex mechanical and transport properties \cite{Pethick1998}. Because of this, it is useful to refer to a possible layer of the crust containing nuclear pasta phases as the mantle, to distinguish it from the expected crystalline region of the inner crust.

The dependence on the symmetry energy of the crust-core transition density and pressure, and hence the crust thickness and extent of the nuclear pasta phases, has been well studied \cite{Kubis2007,Oyamatsu2007,Xu2009,Fattoyev2010,Ducoin2011,Newton2013}. The transition pressure, and hence thickness, mass and moment of inertia of crust, correlates in a non-trivial way with $L$, $K_{\rm sym}$, and to a lesser extent higher-order density expansion parameters of $S(n)$ \cite{Fattoyev2010,Ducoin2011}. The crust composition is also sensitive to the symmetry energy \cite{Newton2013}. The sensitivity to the symmetry energy of other crustal properties important for the modeling of astrophysical phenomena, for example the shear modulus, neutron superfluid gap, thermal and electrical conductivities and entrainment of neutrons by the crustal lattice have yet to be examined explicitly. Nevertheless, we can start to examine the implications for the slope of the symmetry energy $L$ of a number of astrophysical observations of phenomena in which the crust is expected to play a role. We shall review six such observations in this paper.

%====================================== FIGURE 3 ===============================

\begin{figure}
\resizebox{0.45\textwidth}{!}{
\includegraphics{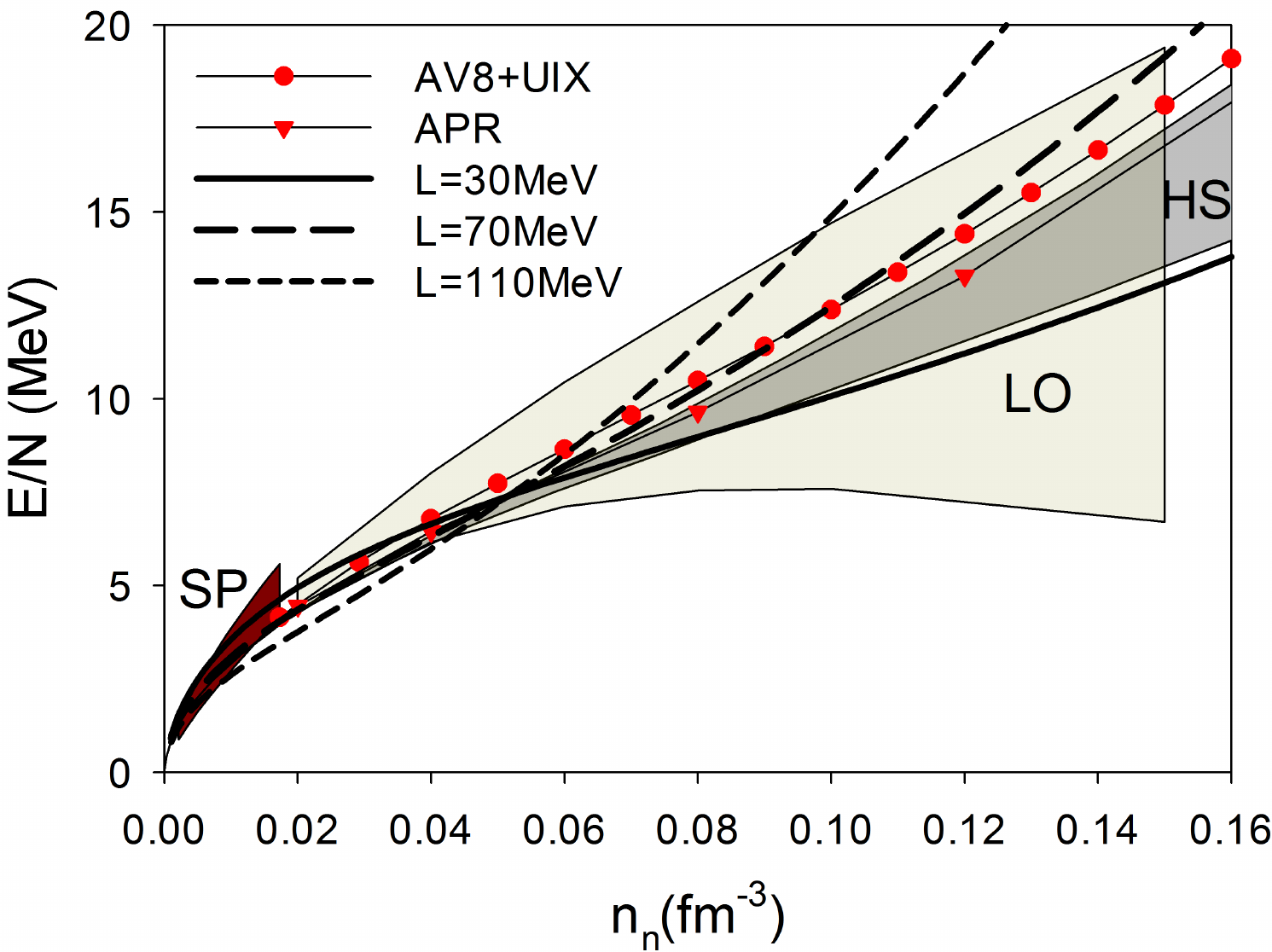}}
\caption{(Color online). Predictions for the energy per neutron of pure neutron matter versus neutron baryon density obtained from our $L$=30,70 and 110 MeV EOSs using the Skyrme model SkIUFSU \cite{Fattoyev2013b,Newton2013d}. These are compared with calculations of Fermi gases in the unitary limit \cite{Schwenk2005} (SP), chiral effective field theory \cite{Hebeler2010} (HS), quantum Monte Carlo calculations using chiral forces at leading order \cite{Gezerlis2013} (LO),  Auxiliary Field Diffusion Monte Carlo using realistic two-nucleon interactions plus phenomenological three-nucleon interactions AV8+UIX \cite{Gandolfi2009,Gandolfi2010}, and the APR EOS \cite{Akmal1998}. Figure adapted from \cite{Newton2013d}.}\label{fig:3}
\end{figure}

%==============================================================================

%====================================== FIGURE 4 ===============================

\begin{figure*}
\begin{center}
\resizebox{0.75\textwidth}{!}{
  \includegraphics{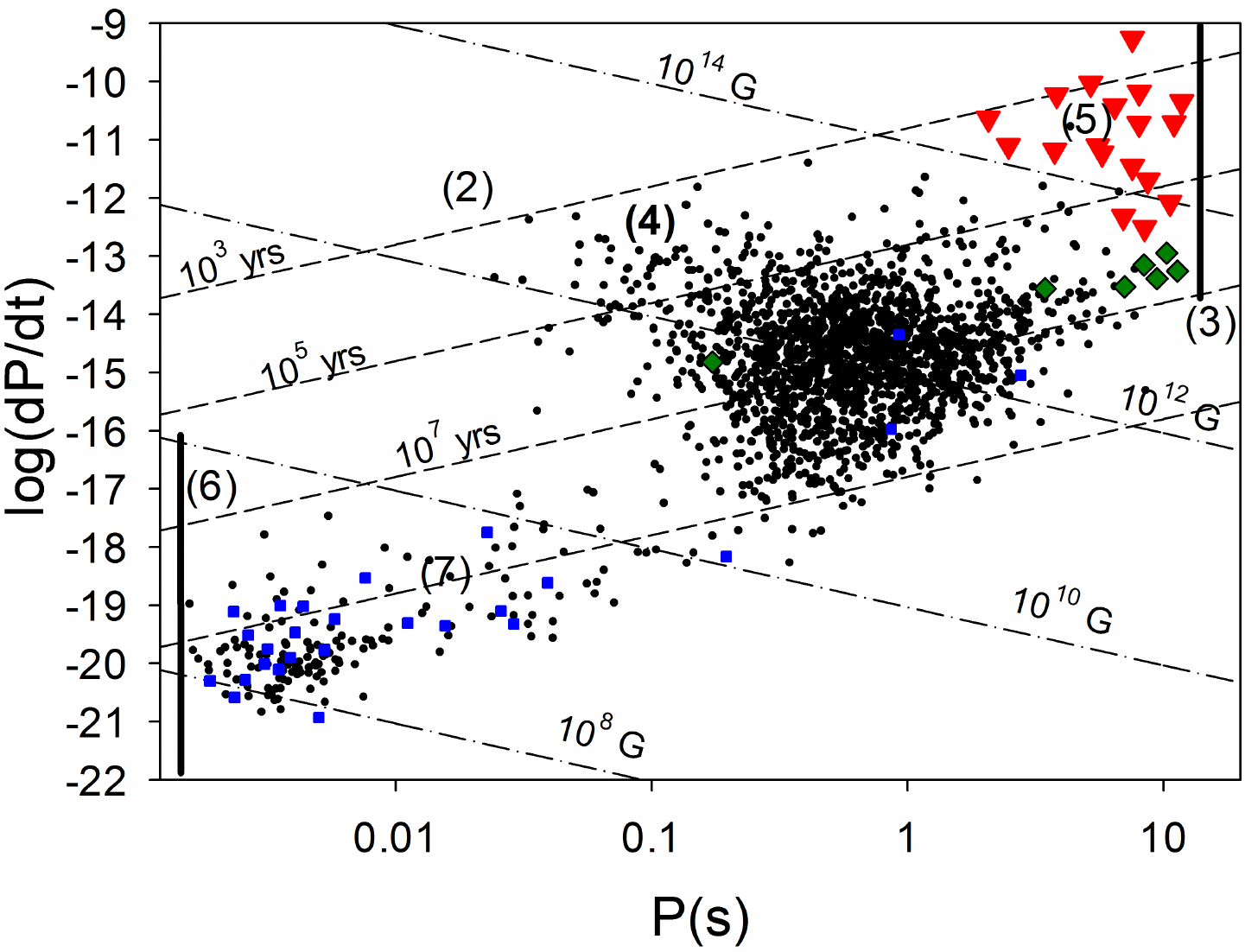}}
\caption{(Color online) Period - period derivative diagram of observed radio and X-ray pulsars. Pulsars observed in the radio band are represented by black dots. Anomalous X-ray pulsars are shown as the red triangles. Isolated neutron stars with thermal X-ray emission are represented by green diamonds, and pulsars known to be members of a binary system are shown as blue squares. Lines of constant characteristic age and magnetic field, inferred from the period and its derivative in a simple dipole spin-down model, are shown by the dashed and days-dotted lines respectively. The six observational probes of the neutron star crust that we discuss in this article are associated with the regions of diagram labelled (2)-(7) in reference to the section in which they are described. Data from \cite{ATNF}.}
\label{fig:4} 
\end{center}
\end{figure*}

%==============================================================================

Some of the studies we review were carried out by our group; let us therefore give a brief outline of the EOSs we use in those studies. We consistently calculate the crust and core EOSs, the crust composition, and the transition densities and pressures between the inner crust and mantle, and crust and core, using the same model for nuclear matter. We use Skyrme and Relativistic Mean-Field (RMF) nuclear matter models. Two parameters of the Skyrme and RMF models affect only isovector properties of matter and can thus systematically adjusted to vary the symmetry energy $J$ and its slope $L$ at saturation density while leaving symmetric nuclear matter properties unchanged \cite{Chen2009,Fattoyev2012a}. We construct single parameter families of neutron star EOSs, parameterized by $L$ \cite{Gearheart2011,Wen2012,Newton2013,Fattoyev2013b}. When adjusting $L$ to obtain a new EOS, we readjust $J$ so that the predicted pure neutron matter (PNM) EOS lies within the robust predictions of \emph{ab initio} calculations at sub-saturation densities; thus the families of EOS contain implicitly a positive correlation between $J$ and $L$ \cite{Fattoyev2012a}. Typically, we vary $L$ in a conservative range  between 25 MeV and 115 MeV. We give examples of the predictions for PNM given by our EOSs for $L=30, 70$ and $110$ MeV together with comparisons with \emph{ab initio} calculations, in Fig.~3. For softer symmetry energies at high densities, the resulting EOS is matched onto two successive polytropes as described in \cite{Wen2012,Steiner2010} in order to match the constraint on the maximum mass of NSs of $M \gtrsim 2M_{\odot}$ \cite{Demorest2010,Antoniadis2013}. Crust models are calculated using a compressible liquid drop model (CLDM) \cite{Newton2013}. The crust and core EOSs used in these works are available from our website \url{http://williamnewton.wordpress.com/ns-eos/}. In Fig.~2 we indicate the ranges of thickness for the inner crust and pasta phases derived from varying the neutron star mass between 1 and 2 $M_{\odot}$ and the symmetry energy slope within the range stated above \cite{NewtonBook2012}. 

Within this model, given the choice of fitting our PNM EOS to robust \emph{ab initio} calculations, the relative thickness, mass and moment of inertia of the crust to the stellar total increases with $L$, whereas the relative thickness and mass of the nuclear pasta layers to the crust total decreases with $L$.

It is useful to place the various astrophysical probes of the crust in the overall context of a neutron star's life cycle. In Fig.~4 we plot the known periods $P$ and period derivatives $\dot{P}$ of neutron stars. Standard neutron stars are believed to be born rapidly rotating and with magnetic fields of order $10^{12}$G, subsequently slowing down due to magnetic braking. Characteristic ages and magnetic fields, derived assuming the model of spin-down due to a dipole magnetic field are shown as the dashed and dash-dot lines respectively. 
Young radio pulsars appear towards the top left of the diagram, and over a timescale of $10^7$ - $10^8$ years, migrate generally downwards and to the right - i.e. towards longer periods and smaller period derivatives (weaker magnetic fields as the field decays over time); eventually the pulsar will no longer be visible. A class of bright X-ray pulsars, so-called anomalous X-ray pulsars (AXPs), are believed to be powered by extremely strong magnetic fields; such neutron stars are thus referred to as magnetars. They appear in the top right corner of the diagram (the red triangles). Such stars can also undergo regular bursts in X-ray and soft gamma-ray, as a result of magnetic field rearrangement and crust cracking; these objects are referred to a soft gamma-ray repeaters (SGRs). If a member of a binary system, in later life the neutron star can accrete matter off its companion star. The resulting transfer of angular momentum spins the star up, often to millisecond periods, and the star may become visible again in the bottom left corner of the diagram, its emission now powered by accretion and thermonuclear bursts on its surface. 

A neutron star can therefore experience a rather rich life and exhibit a wide variety of observable phenomena; see \cite{Kaspi2010,Harding2013} for more detailed reviews of the neutron star `zoo'. Substantial recent work has attempted to unite the various observational classes of neutron stars under a single evolutionary model \cite{Vigano2013,Ho2013}. Such diversity of observational phenomena is welcome as we seek a variety of independent probes of neutron star physics to test our models against. The regions of the period-period derivative diagram associated with the observables we shall be reviewing in this paper are labelled (2)-(7) in Fig.~4 for the section they appear in. They are as follows. Section (2): the cooling of the neutron star in Cassiopeia A (this object does not have a measured $P$ or $\dot{P}$, so its label is only indicative). Section (3): upper limit of the spin period of X-ray pulsars (appearing as the bar in the top-right of the diagram). Section (4): Glitches from the Vela pulsar. Section (5): Quasi-periodic oscillations in the light curves of giant flares from SGRs. Section (6): lower limit of the spin period of millisecond pulsars, depicted as the bar in the lower left corner of the diagram. Section (7): electromagnetic pre-cursors to soft gamma-ray bursts, thought to result from the merger of two neutron stars. In each of the sections we will discuss the inferred constraints on $L$, and make clear in a series of concluding caveats the main limitations of models used to obtain those constraints and the interpretations of observations required to make them. In section~8, we shall summarize constraints on $L$ and deliver some concluding comments. 

%%%%%%%%%%%%%%%%%%%%%%%%%%%%
%
% CAS A
%
%%%%%%%%%%%%%%%%%%%%%%%%%%%%

\section{The cooling of the neutron star in the Cassiopeia A supernova remnant}
\label{sec:2}

The Cassiopeia supernova is estimated to have occurred, and the resulting neutron star born, in the year 1680 $\pm$ 20 \cite{Fesen2006}. In 2009, thermal X-ray emission from the Cas A neutron star (CANS) was fit utilizing a Carbon atmosphere model \cite{Ho2009}, giving an average surface temperature $\langle T_{\rm eff} \rangle \approx 2.1\times10^6$K as measured at the stellar surface. We note that the Carbon atmosphere composition is preferred on the basis that it gives an emitting area of order the neutron star size. Further analysis of data taken from the Chandra X-ray telescope, taken over the previous decade, indicates a rapid decrease in surface temperature by $\approx 4\%$ \cite{Heinke2010}. A recent detailed analysis of the Chandra data, including all X-ray detectors and modes, provides a more conservative 2-5.5$\%$ temperature decline over the same time interval \cite{Elshamouty2013};  another recent analysis concludes, however, that the current data is consistent with there being no cooling from the CANS \cite{Posselt2013}. Even a 2$\%$ decline would constitute a rapid cooling phase. A definitive measurement is difficult due to surrounding bright and variable supernova remnant, and further data from Chandra over the next decade or more is required to settle the issue.

Within the minimal cooling paradigm (MCP) \cite{Page2004}, the rapid cooling of the CANS is interpreted as the result of enhanced neutrino emission from neutron Cooper pair breaking and formation (PBF) as the core neutrons transition to a superfluid phase, thus providing the first evidence for stellar superfluidity (\cite{Shternin2011,Page2011}; however, alternative explanations involving in-medium modifications to the modified Urca (MU) neutrino emission process in the core, neutrino emission processes in exotic phases such as quark matter, and heating of the crust due to dissipation of small-scale magnetic fields in the crust  have been proposed \cite{Blaschke2012,Sedrakian2013,Bonanno2013}. Core neutrons (protons) form Cooper pairs in the $^3$P$_2$ ($^1$S$_0$) channel; the critical temperatures $T_c$ for the onset of superfluidity are strongly density dependent, and suffer significant theoretical uncertainty.
 
The maximum value for the $^3$P$_2$ critical temperature determines the age at which the PBF cooling phase is entered. $T_c$ can be tuned so that the PBF cooling curves pass through the observed temperature of the CANS at the age of  $\approx335$ years. Using the APR EOS \cite{Akmal1998}, the PBF cooling epoch best fit the data with a critical temperature of  $T_c=5-9\times10^8$K \cite{Page2011,Shternin2011}. The \emph{rate} of cooling during the PBF phase is determined by the core temperature at the onset of the PBF cooling phase, $T_{\rm PBF}$. A higher $T_{\rm PBF}$ leads to a steeper cooling rate. The inferred $\approx$4\% temperature decline is relatively steep compared to the predictions of theoretical cooling trajectories, and it was found that it could be reproduced only if the protons throughout core of the star had already undergone a transition to superconductivity, thus suppressing the MU process in the core and raising the core temperature.

%====================================== FIGURE 5 ===============================

\begin{figure}[t]
\begin{center}
\includegraphics[width=18pc]{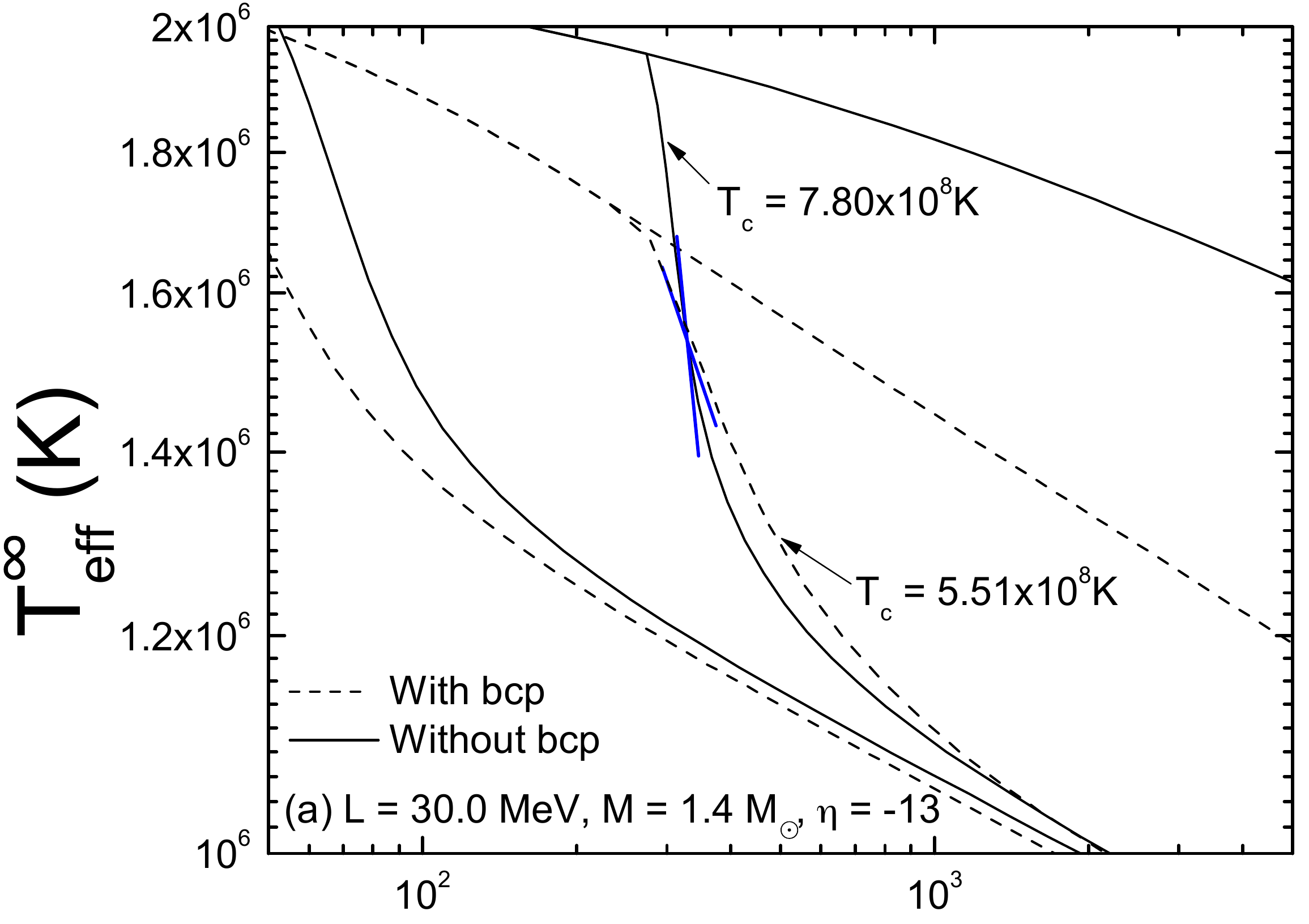}
\includegraphics[width=18pc]{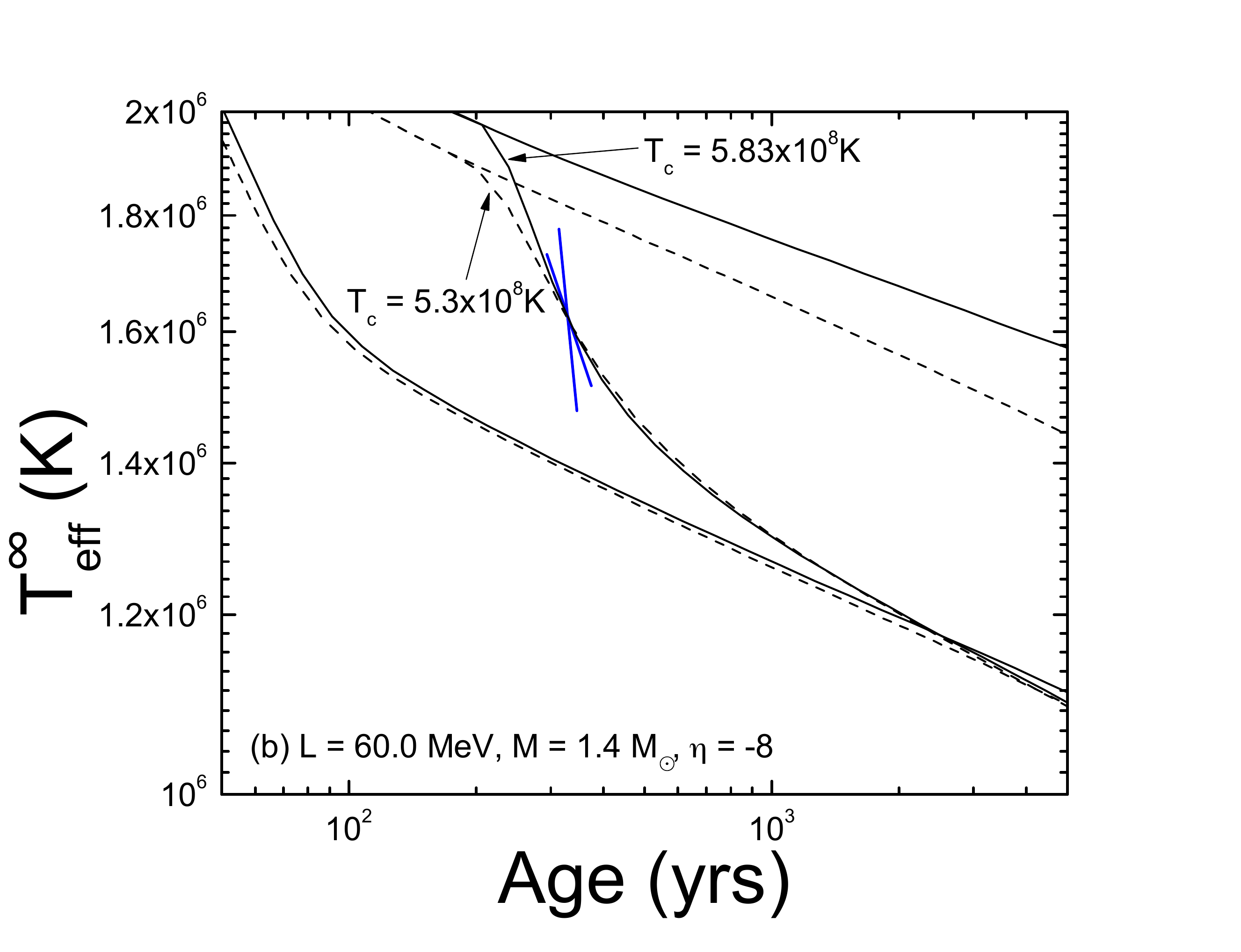}
\caption{Two pairs of cooling windows, together with the cooling curves which pass through the average value of the measured CANS temperature and the corresponding value of $T_{\rm cn}^{\rm max}$. Each plot shows the window with BCPs inactive (solid lines) and active (dashed lines) for combinations ($L$(MeV), $M/M_{\odot}$, $\eta$) of (30, 1.4, -13) (a) and (60, 1.4, -8) (b) ). The largest and smallest cooling rates estimated from observations are indicated by the straight lines passing through the average CANS temperature. Taken from \cite{Newton2013d}.}
\label{fig:5} 
\end{center}
\end{figure}

%===============================================================================

In the MCP, three additional parameters affect the cooling trajectories of the NSs \cite{Page2004}: 

\noindent (i) The EOS of nuclear matter (NM).

\noindent (ii) The mass of light elements in the atmosphere $\Delta M_{\rm light}$ parameterized here as $\eta = \log \Delta M_{\rm light}$. The thermal spectrum of CANS can be fit using light element masses $-13\lesssim \eta \lesssim -8$  \cite{Yakovlev2011}. More light elements means higher thermal conductivity and lower core temperature for a given $T_{\rm eff}$.

\noindent (iii) The mass of CANS, constrained by atmosphere modeling to $\approx$ 1.25 - 2$M_\odot$ with a most likely value of 1.65$M_\odot$ \cite{Yakovlev2011}.

In the original analysis using the PBF cooling paradigm, this parameter space was largely unexplored; especially, the EOS dependence of the results was not examined. Parameterizing the EOS by the slope of the symmetry energy $L$, in \cite{Newton2013d} we examined the sensitivity of the results to the slope of the symmetry energy via the core EOS.

There are additional neutrino cooling processes that have been postulated to operate in the pasta layers of the neutron star crust; if such processes are important, they will also be sensitive to $L$ via the dependence of the volume of the pasta in the crust on $L$. Due to momentum conservation from scattering of neutrons off bubble phases of pasta, the direct Urca (DU) neutrino cooling process might occur in the bubble phases due to neutron scattering \cite{Gusakov2004}. Also, neutrino and anti-neutrino pairs can be emitted from nucleon-pasta scattering \cite{Leinson1993}. We shall refer to these collectively as bubble cooling processes (BCPs).  The neutrino luminosity of BCPs goes as $L_{\rm \nu}^{BCP} \sim 10^{40} T_9^6$ where $T_9 = T_{\rm core}/10^9$K. The neutrino luminosity from the modified Urca process, assuming it to be uninhibited by superconductivity, goes as $L_{\rm \nu}^{MU} \sim 10^{40} T_9^8$. Therefore, at core temperatures of order the critical temperature for the onset of neutron superfluidity $T_9 = 0.5 - 1$, these neutrino emission luminosities are comparable. Therefore, even if superconductivity suppresses the MU process, BCPs might provide additional cooling which will lessen the calculated cooling rate during the PBF cooling phase, therefore making a match to the inferred cooling rate more difficult. In \cite{Newton2013d} we tested the effect of these crustal processes on the inferred range of parameters for which the PBF cooling scenario was consistent with observation, and thereby obtain constraints on $L$ through both the global cooling and the crustal cooling.

To illustrate the effect of the BCPs, in Fig.~5 we show sets of cooling trajectories for $1.4M_{\odot}$ stars with two different EOSs ($L$=30 MeV (a) and $L$=60 MeV (b)) and envelope compositions ($\eta = -13, -8$). The solid lines show the cooling trajectories without the inclusion of BCPs, and the dashed lines the BCP-modified trajectories. The upper and lower trajectories correspond to values of the critical temperature $T_{\rm cn}^{\rm max}$ of 0K (no $^3$P$_2$ neutron pairing) and $T_{\rm cn}^{\rm max}$ = $10^9$K (lower trajectories); each pair of trajectories $T_{\rm cn}^{\rm max}$ = 0K, $10^9$K, forms a cooling window inside which the observed temperature must fall. The upper and lower bounds on the cooling rate inferred from the X-ray data are indicated by the two straight lines, which intersect at the average observed surface temperature at Earth $\langle T^{\infty}_{\rm eff} \rangle$, a temperature which takes into account the gravitational redshift of the emission.The trajectories which best fit $\langle T^{\infty}_{\rm eff} \rangle$ are shown with the corresponding best-fit value of the critical temperature $T_{\rm c}$. 

One can see that with the BCPs active, the temperature at the onset of the PBF phase is smaller and the cooling trajectory shallower. The effect becomes less pronounced for higher value of $L$ as in those cases, the pasta phases occupy a smaller volume fraction of the crust.
At $L=30$ MeV, both cooling trajectories fit the inferred cooling rate (i.e. fall between the two straight lines); for $L=60$ MeV, neither trajectory is consistent with the inferred cooling rate. At intermediate values of $L$, only the trajectory without BCPs will be consistent with the inferred cooling rate.  Varying the set of parameters $L, \eta, M$, BCPs on/off, we find the following variations lead to a lower core temperature and hence a shallower cooling rate which is harder to match to the data:

\begin{itemize}
        \item{Inclusion of BCPs narrows the cooling window.}
        \item{Increasing the mass of light elements in the envelope $\Delta M_{\rm light}$ (increasing $\eta$) leads to a lower core temperature for a given surface temperature $T_{\rm eff}^{\infty}$.}
        \item{As $M$ increases, the central stellar density increases and the fraction of the core in which the protons are superconducting decreases, making the modified Urca (MU) process more efficient and the star's core cooler.}
        \item{Increasing $L$ increases the radius $R$, requiring a lower $T_{\rm eff}$ and hence core temperature to produce the same stellar luminosity.}
\end{itemize}

Even the 2\% temperature decline over the time period 2000-2009 is relatively rapid; the data thus favors a relatively high core temperature and thus smaller values of $L$ (smaller radii), smaller stellar masses $M$, smaller $\Delta M_{\rm light}$ (smaller $\eta$) and no BCPs. With active BCPs the best fit cooling curve in the PBF phase is significantly less steep than without BCPs, and matches only the shallowest inferred cooling rate, and then only for the lowest values of $L$. As $L$ increases beyond $50-60$ MeV, depending on mass, the curves become too shallow even with no BCPs operating. For each set of parameters $\eta, M$, BCPs on/off, we can find an upper limit on the value of $L$ consistent with the inferred cooling rate.
   
Accepting the MCP cooling model and the accuracy of X-ray measurements and interpretation, we obtain the following constraints from fitting the average CANS temperature and the cooling rate simultaneously:
\begin{itemize}
\item{For BCPs inactive, $L\lesssim 70$ MeV ($35\lesssim L\lesssim 55$ MeV) for $1.25 M_{\odot} < M < 1.8M_{\odot}$ ($M=1.6 M_{\odot}$).}
\item{For BCPs active, $L\lesssim 45$ MeV ($35\lesssim L\lesssim 45$ MeV) for $1.25 M_{\odot} < M < 1.8M_{\odot}$ ($M=1.6M_{\odot}$).}\\
\end{itemize}

\noindent \textbf{\emph{Constraint:}} $L\lesssim 45$ MeV (with pasta cooling processes), $L\lesssim 70$ MeV (without pasta cooling processes)

\noindent \textbf{\emph{Caveats:}}. The Carbon atmosphere model is preferred because it gives X-ray emitting areas of order NS radii; other atmosphere models would shift $\langle T_{\rm eff}^{\infty} \rangle$ (thus the inferred range of $L$), but not the cooling rate. Enhanced superfluidity in the crust is not ruled out and would exponentially suppress BCPs, making our constraints closer to those obtained with no BCPs active. Medium modifications to neutrino cooling processes and cooling processes from exotic components could substantially alter the conclusions reached in the minimal cooling scenario. More recent X-ray data from the CANS is consistent with there being no cooling at all. 

%%%%%%%%%%%%%%%%%%%%%%%%%%%%%
%
% SPIN DOWN EVOLUTION 1: LIMITING SPIN PERIOD
%
%%%%%%%%%%%%%%%%%%%%%%%%%%%%%

\section{Upper limit to the observed spin periods of young pulsars}
\label{sec:3}

After their birth as rapid-spinning objects, the dynamical evolution of most pulsars is dominated by a spin-down epoch of $\sim 10^7 - 10^8$ years (the young - middle ages of their lives), caused in large part by the torque applied by their large magnetic fields. How long a period can a pulsar spin down to during this phase of evolution? To answer this question, we must look to observations of \emph{X-ray} pulsars rather than radio pulsars, because they will be free of observational biases against long-period pulsars that are apparent in radio observations.

When one looks at the periods of young to middle aged X-ray pulsars, the highest periods we observe are of order 10s (see Fig.~1). Given the lack of selection effects in the X-ray band, this can be taken as a physical upper limit to the period of a neutron star with an age $\lesssim 10^7$ yrs. However, many of the longer period objects, $P\sim 1-10$s between $10^3$ and $10^6$ years old, also exhibit large magnetic fields $\sim 10^{13} - 10^{14}$G and above, inferred from the spin-down rates. With such fields, these stars should spin down to periods of 100s and above on timescales of $\rm 10^4$ yrs, still within the detectable range of X-ray measurements. Why do we see no such long period pulsars with high magnetic fields?

This inconsistency can be resolved by taking into consideration the decay of the magnetic field. A decay of magnetar strength fields to strengths of $\lesssim 10^{12}$ G on timescales of order $10^6$yrs would inhibit the spin-down of the star sufficiently that its period would not exceed $\sim10$s \cite{Pons2013}. There are two main mechanisms for magnetic field decay in the neutron star crust \cite{Goldreich1992,Cumming2004,Pons2007}. At early times, Hall drift dominates the field decay, replaced by Ohmic decay at ages of $\sim 10^5$yrs which are the ages of interest in this particular problem. At ages beyond $\sim 10^5$yrs, crust temperatures are cool enough that the main contribution to the electrical resistivity of the crust is through electron scattering off of impurities. A measure of the impurity content of the crust is given by the parameter

\be
Q_{\rm imp} =  \sum x_i [ Z_i - \langle Z \rangle]^2
\ee

\noindent where the sum is over the ionic species $i$ present in the crust, $x_i$ is the fraction of species $i$ present, $Z_i$ is the proton number of the species and $\langle Z \rangle$ is the average proton number of ions in the crust at that particular density. Generally, $Q_{\rm imp} \ll 1$ indicates a relatively pure, homogeneous lattice, while $Q_{\rm imp} > 1$ indicates an impure, heterogenous lattice. Ref. \cite{Pons2013} performed consistent magneto-thermal evolutions for high magnetic field neutron stars, and showed that Ohmic decay can reduce the magnetic field sufficiently in crust to explain the observed X-ray pulsar period distribution if there is a highly resistive layer $Q_{\rm imp} \approx 20-100$ at the base of the inner crust (above densities of $6\times10^{13}$ g cm$^{-3}$), consistent with the existence of a layer of nuclear pasta which is expected to be highly disordered and amorphous \cite{Magierski2002}. Evidence for a significant pasta layer in the crust sets loose constraints on the slope of the symmetry energy $L\lesssim 80$ MeV \cite{Oyamatsu2007,Newton2013}.

Although the observed period distribution of X-ray pulsars is consistent with a significant pasta layer, it is not the only explanation. The structure of the inner crust of the neutron star is still quite uncertain theoretically. State-of-the-art semi-classical molecular dynamics simulations indicate the crust is very pure at lower densities $Q_{\rm imp} \ll 1$ \cite{Hughto2011}. Thermodynamic arguments, on the other hand, point towards the inner crust being quite disordered throughout its extent $Q_{\rm imp} \sim 20$, also consistent with the period distribution \cite{Jones2004a,Jones2004b}. Recently, it was shown that a bcc lattice is rendered unstable because of an effective attractive interaction between ions induced by the free neutrons \cite{Kobyakov2013}, and that as a result the crust could assume a more exotic crystal structure; the effect on the resistivity is an open question. 

Fortunately, there is also another observational probe of the crustal structure. Observations of the cooling of older neutrons stars whose crusts are in the process of thermal relaxation after a period of heating caused by accretion induced nuclear burning are consistent with a relatively pure inner crust, with $Q_{\rm imp} \sim 1$ \cite{Brown2009}. These observations are less sensitive to the very deepest layers of the crust where the pasta phases are expected; consistency with the period distribution of X-ray pulsars supports the existence of a highly resistive pasta layer. However, the crusts in such neutron stars may be completely or partially recycled, meaning that periods of accretion from a binary companion may have replaced some or all of the original crust, with the original material having sunk into the core. Recent observations of cooling of a younger neutron star IGR J17480-2446 after accretion induced heating show that the star is hotter than models using recycled crusts appropriate for older neutron stars predict \cite{Cackett2010,Degenaar2013}. The suggestion is that the original, cold catalyzed neutron star crust is more impure than the replacement.

Despite the theoretical uncertainty, the observations are already probing the structure of the inner crust in ways that give strong hope for tighter constraints on the crustal structure and thus on the symmetry energy in the future.\\

\noindent \textbf{\emph{Constraint:}} $L\lesssim 80$ MeV.

\noindent \textbf{\emph{Caveats:}} Impurity level of the pasta phases yet to be rigorously modeled. Some observational and theoretical evidence that the inner crust is quite impure throughout, removing the need for a highly impure pasta layer.

%%%%%%%%%%%%%%%%%%%%%%%%%%%%%%
%
% SPIN DOWN EVOLUTION II: GLITCHES
%
%%%%%%%%%%%%%%%%%%%%%%%%%%%%%%

\section{Glitches in the Vela pulsar}
\label{sec:4}

Young pulsars $\lesssim10^{-7}$yr are frequently observed to have their spin-down evolution interrupted by sudden increases in their spin frequency $\nu$, events known as glitches. A wide range of glitch sizes $\frac{\Delta \nu}{\nu} \sim 10^{-11}$ to $10^{-5}$ and recovery timescales and behaviors have been reported \cite{Espinoza2011g}. 

In order to probe the mechanism responsible for glitches, one can attempt a population synthesis of all glitches observed to date and attempt to develop models that account for the variability in the population statistically \cite{Melatos2007g} or one can examine a particular pulsar from which a large number of glitches have been observed, and attempt to account for the details of individual glitches. We recap here recent progress on the latter front examining the behavior of glitches from the Vela pulsar.

21 giant glitches have been observed (giant meaning $\frac{\Delta \nu}{\nu_{0}} \gtrsim 10^{-6}$) from the Vela pulsar, and the recurrence times appear quasi-periodically distributed with an inter-glitch timescale of $\approx 3$yrs \cite{Melatos2007g}.  The timescale on which the frequency increase occurs (the glitch rise time) has been constrained to $\lesssim 40$s  \cite{Dodson2002g}.

The leading class of glitch models postulate two components of the neutron star interior that for the majority of the time are dynamically decoupled, with one component coupled strongly to the magnetosphere and hence spinning down due to magnetic braking (this is the component whose spin frequency we observe), while the other component spins at a constant rate and hence accumulates an excess of angular momentum relative to the former component (this component is often referred to as angular momentum ``reservoir''). Occasionally the two components strongly couple, and the reservoir transfers angular momentum to the component coupled to the magnetosphere and we observe a sudden frequency increase.

The version of this model that has gained most traction posits that the angular momentum reservoir consists of the superfluid neutron immersing the nuclei and pasta in the crust \cite{Anderson1975g,Alpar1977g}. The superfluid forms a two-dimensional array of vortices whose area density is proportional to the spin frequency; for the superfluid to spin down with the rest of the star, the vortices must move radially outwards from the rotation axis. In the core, the vortices are able to do this; however, neutron superfluid vortices in the crust interact with the nuclei there and become pinned in the lattice \cite{Avogadro2008g,Pizzochero1997g}, preventing further outward motion and thus becoming decoupled from the rest of the star \cite{Alpar1984g,Alpar1977g,Anderson1982g,Pines1980g}. As the difference between the angular velocity of the two components increases, the Magnus force on the inner crust neutron vortices grows until, when the angular momentum lag exceeds a threshold value, it overcomes the effective pinning force and vortices unpin as a whole, coupling to the charged component and spinning it up.

As a model for pulsar glitches, there are several problems with the current paradigm which are actively being worked on: there are problems with how well the model predicts the post-glitch recovery of the spin frequency to its pre-glitch behavior, and the details of how the glitch is initiated. However, at a more fundamental level, the model should be consistent with the observed inter-glitch timescales and the glitch sizes. Here we focus on the latter in relation to the Vela pulsar, and examine the fundamental question of whether the crust-initiated glitch model can account for glitches of the size observed in the Vela pulsar, which sensitively depends on the EOS and, particularly, the symmetry energy at saturation density.

%====================================== FIGURE 6 ===============================

\begin{figure}\label{fig:6}
\begin{center}
\includegraphics[width=8cm,height=8cm]{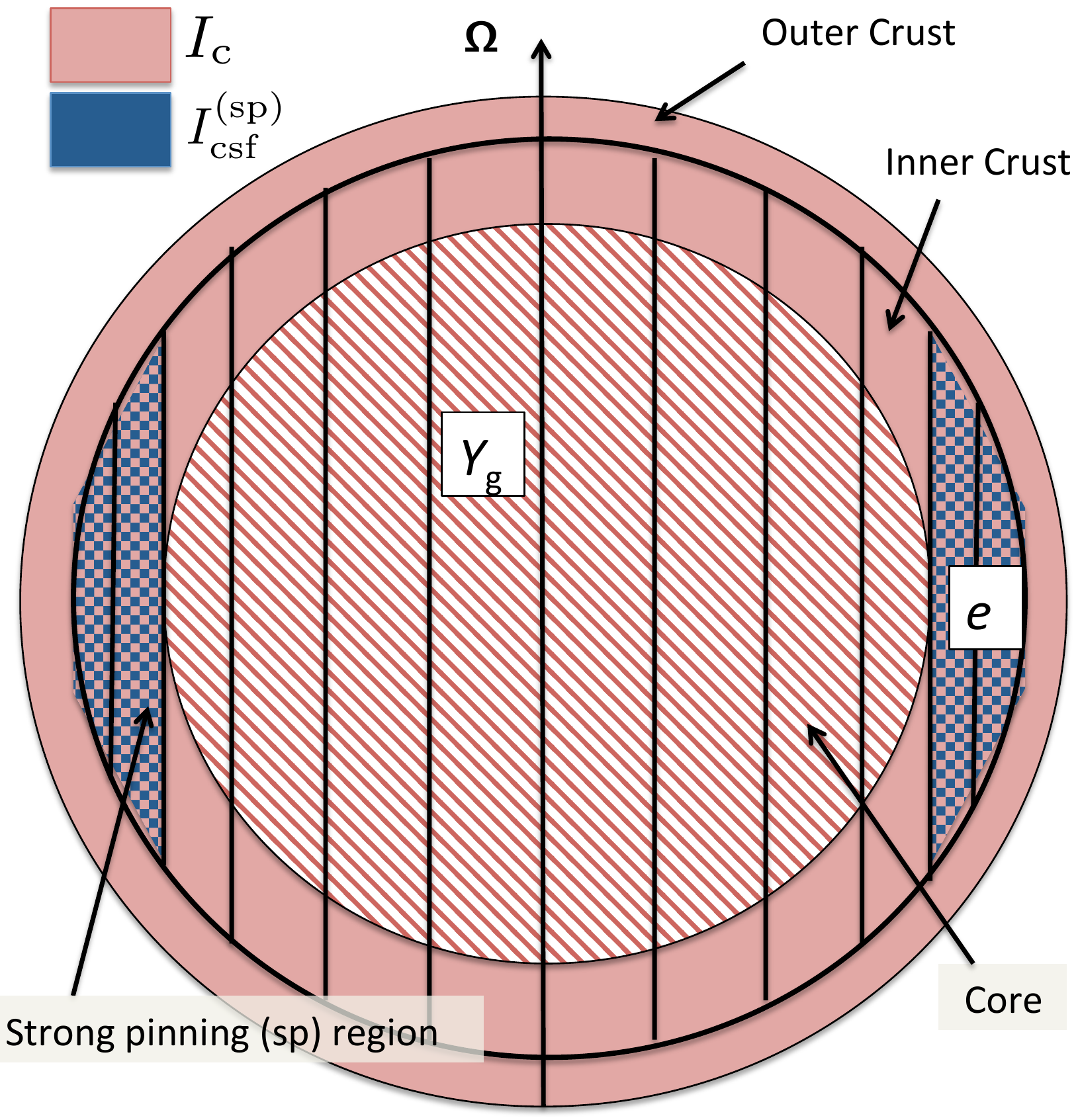}
\caption{(Color online). Illustration of the ``snowplough'' glitch mechanism. Neutron vortices are represented by the straight black lines aligned with the rotation axis $\Omega$. To be pinned effectively they must be fully immersed in the crustal lattice, which only occurs in the strong pinning region of the inner crust. The red regions in the crust and core represent the contributions to the moment of inertia of the star from the charged components of the star and the core neutrons that are strongly coupled to them (the fraction of which is given by $Y_{\rm g}$.) The fraction of free crustal superfluid neutrons - that is, neutrons not entrained by the crustal lattice - is depicted in blue; the strength of entrainment in the crust is parameterized by $e$.}
\end{center}
\end{figure}

%==============================================================================

Let us denote the moment of inertia of the angular momentum reservoir $\Delta I$, the moment of inertia of that part of the star that couples strongly to the angular momentum reservoir on timescales of order the glitch rise time ($\lesssim 40s$ for Vela) by $I$ (which has an upper limit of the total moment of the star by $I_{\rm tot}$). Define the parameter $G$ as \cite{Link1999g}

\be
G \equiv \frac{\Delta I}{I} \geq \frac{\bar{\Omega}}{|\dot{\Omega}|} \mathcal{A}
\ee

\noindent where $\mathcal{A}$ is the glitch activity parameter of the pulsar, that is the slope of the straight line fit to a plot of the cumulative relative glitch size over time \cite{Link1999g}, $\bar{\Omega}$ is the average spin frequency over the period of time during which the glitches occurred and  $\dot{\Omega}$ is the steady-state spin down rate between glitches. Then analysis of the rate of spin-up of the Vela pulsar caused by glitches gives us an observational constraint of $\Delta I/I \gtrsim 1.6 \%$ \cite{Espinoza2011g,Link1999g}.

It was shown that, taking $I = I_{\rm tot}$ and $\Delta I$ as the moment of inertia of the whole crust as an upper limit to the moment of inertia of the crustal superfluid neutrons $I_{\rm csf}$, many realistic neutron star EOSs can satisfy $\Delta I/I \gtrsim 1.6 \%$, and moreover that the dependence of the crustal moment of inertia on the EOS potentially allows constraints to be set on the symmetry energy \cite{Lorenz1992g,Link1999g,Fattoyev2010}. In the initial studies, it appeared that all but the softest EOSs predicted maximum glitch sizes consistent with Vela. However, the effect of entrainment of superfluid neutrons by the crustal nuclei was being neglected. Bragg scattering by neutrons off of nuclei, analogous to the similar effect for electrons in metals which gives rise to the band structure of the single particle energy levels, couples a certain fraction of the neutrons to the crustal lattice, thus reducing the fraction of superfluid neutrons decoupled from the crust by a factor of, on average, $\approx 5$ (that is $\Delta I \sim 0.2 I_{\rm csf}$). This would appear to reduce $\Delta I /I$ too much to explain the sizes observed in the Vela pulsar \cite{Chamel2012ga,Chamel2012gb,Andersson2012g}. However, these studies assume that the crust-core coupling at the time of glitch is sufficiently strong that the whole star spins up with the crust: $I \sim I_{\rm tot}$. The crust-core coupling strength is still rather uncertain, however, and estimates of the  coupling timescales due to interactions of neutron vortices and type-I or type-II superconducting protons (\cite{Alpar1988g,Sedrakian2004g,Andersson2005g,Jones2006g,Babaev2009g,Link2012g}) are consistent with the possibility that, on the glitch rise timescale $<40$s \cite{Dodson2002g}, only a small fraction of the core neutrons are coupled to the crust and therefore $I \ll I_{\rm tot}$. This would significantly reduce the ratio $\Delta I /I$ allowing it once again to satisfy the lower bound of 1.6\% even allowing for crustal neutron entrainment at the level suggested by theory.

The picture is still not complete. A recent hydrodynamical model of glitches \cite{Haskell2011g,Pizzochero2011g,Seveso2012g}, referred to by its authors as the ``snowplough'' model, shows that vortices are pinned in the crust only in the region in which they are totally immersed, a ``napkin ring'' shaped annulus of in the inner crust which contains only $\sim 10$\% the mass of the whole crust. This is because vortices that thread the core, and are only pinned at in the crust at their ends, can creep outwards at a rate which tracks the overall spin evolution of the crust, with only a very small lag in angular velocity. This implies that the moment of inertia of the crust neutrons available to participate in the glitch event is further reduced by a factor of about 10 $\Delta I \to 0.1 \Delta I$. In the snowplough model, $I \ll I_{\rm tot}$ as outlined above, and so the ratio $\Delta I / I$ is still able to account for the observed Vela glitch activity for selected EoSs \cite{Seveso2012g}, but the effect of crustal entrainment is not included. 

Recently, we performed estimates of the efficacy of glitch models to reproduce the observed glitch sizes in the Vela pulsar including \emph{all} the effects above that modify the critical ratio $\Delta I/I$ which we outlined above. To summarize, (i) the reduction of $\Delta I$ by a factor of $\sim 5$ due to entrainment of crustal superfluid neutrons, (ii) the reduction of $\Delta I$ by a factor of $\sim 10$ due to the localization of the strong pinning region to an annulus of the inner crust, and (iii) the reduction of the moment of inertia of the core that is coupled to the crust at the time of the glitch $I$ due to the fact that only a fraction of core neutrons $Y_{\rm g}$ are coupled strongly to the charged component of the star then \cite{Hooker2013}. We explored the EOS and crust-composition dependence by using a family of SkIUFSU Skyrme crust and core EOSs \cite{Fattoyev2012a}, including the consistently calculated neutron fraction in the crust and core. We now summarize the specific model and results.

This model is illustrated in Fig. 6. The component of the star that effectively acts as the angular momentum reservoir for the glitch - the fraction of crustal superfluid neutrons in the strong pinning region that are not entrained by the crustal lattice, $1-e$, are shown in the blue (dark colored) region and have a moment of inertia $I^{\rm (sp)}_{\rm csf}$. The charged component of the star includes the crust lattice plus the protons in the core and some fraction of core superfluid neutrons $Y_{\rm g}$ that couple to the protons on timescales shorter than the glitch rise time are shown as the red (lighter colored) region, and have a total moment of inertia given by $I_{\rm c}$. The former spins up the latter in the glitch event.  

%====================================== FIGURE 7 ===============================

\begin{figure*}\label{fig:7}
\begin{center}
\includegraphics[width=6cm,height=5cm]{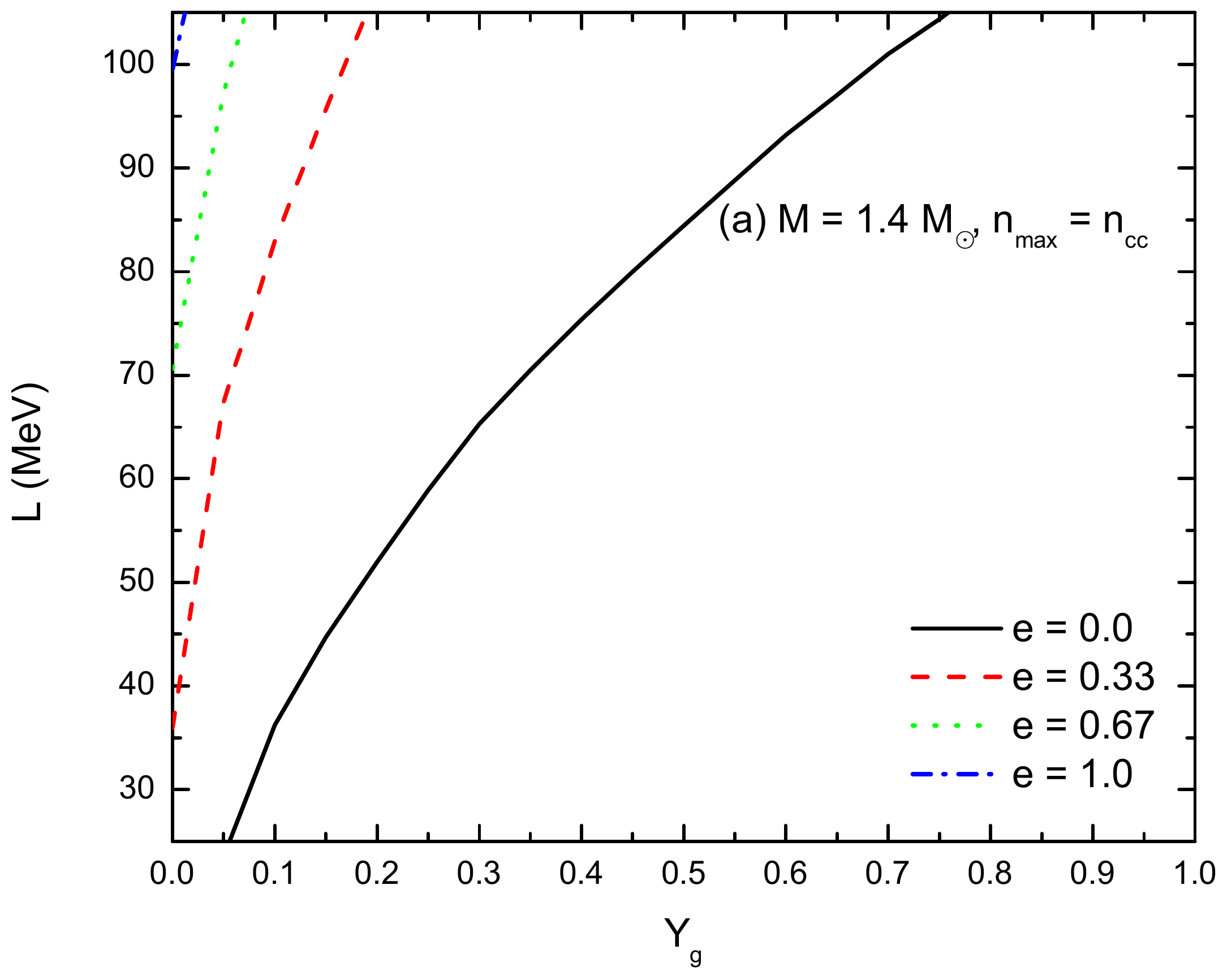}\includegraphics[width=6cm,height=5cm]{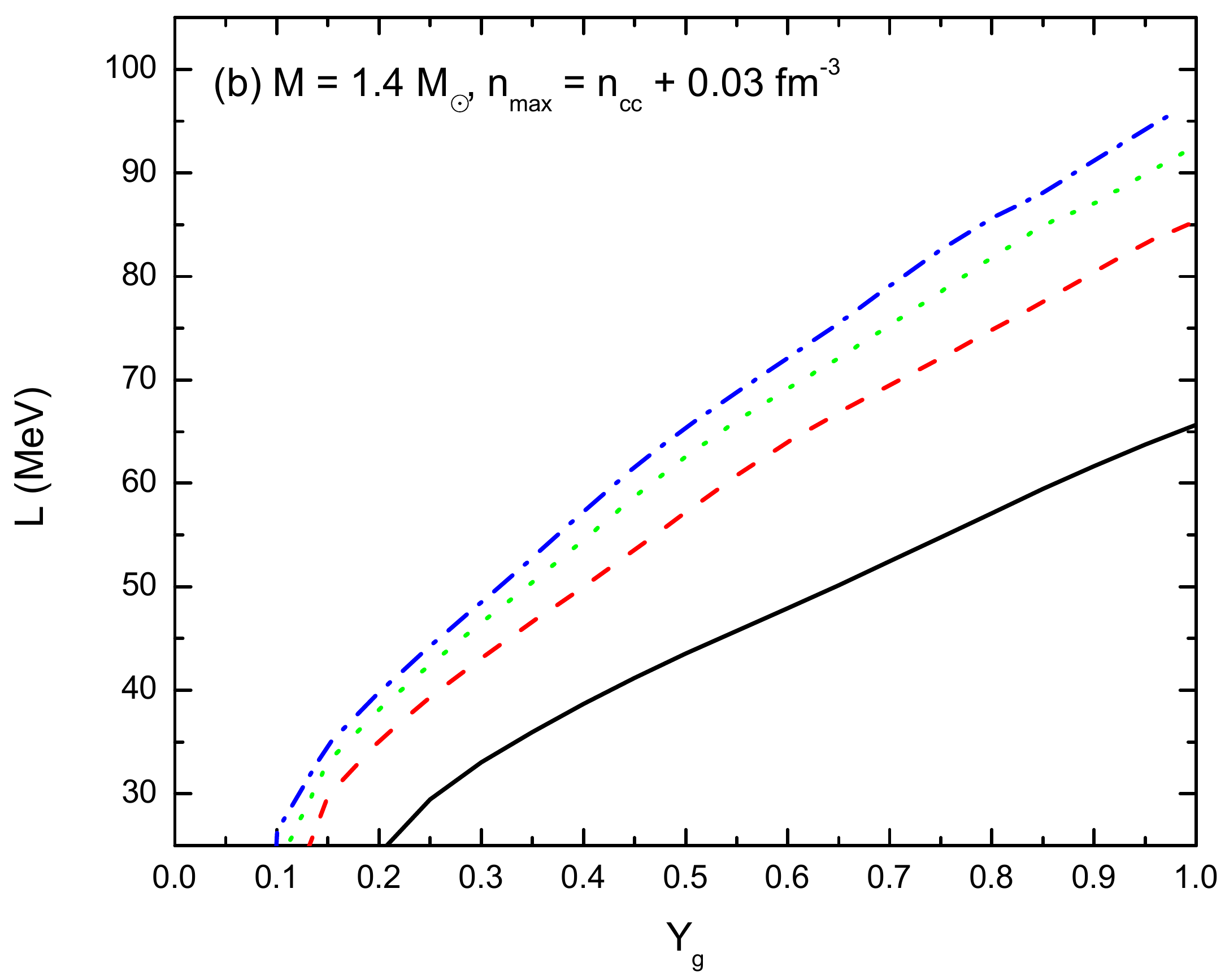}\includegraphics[width=6cm,height=5cm]{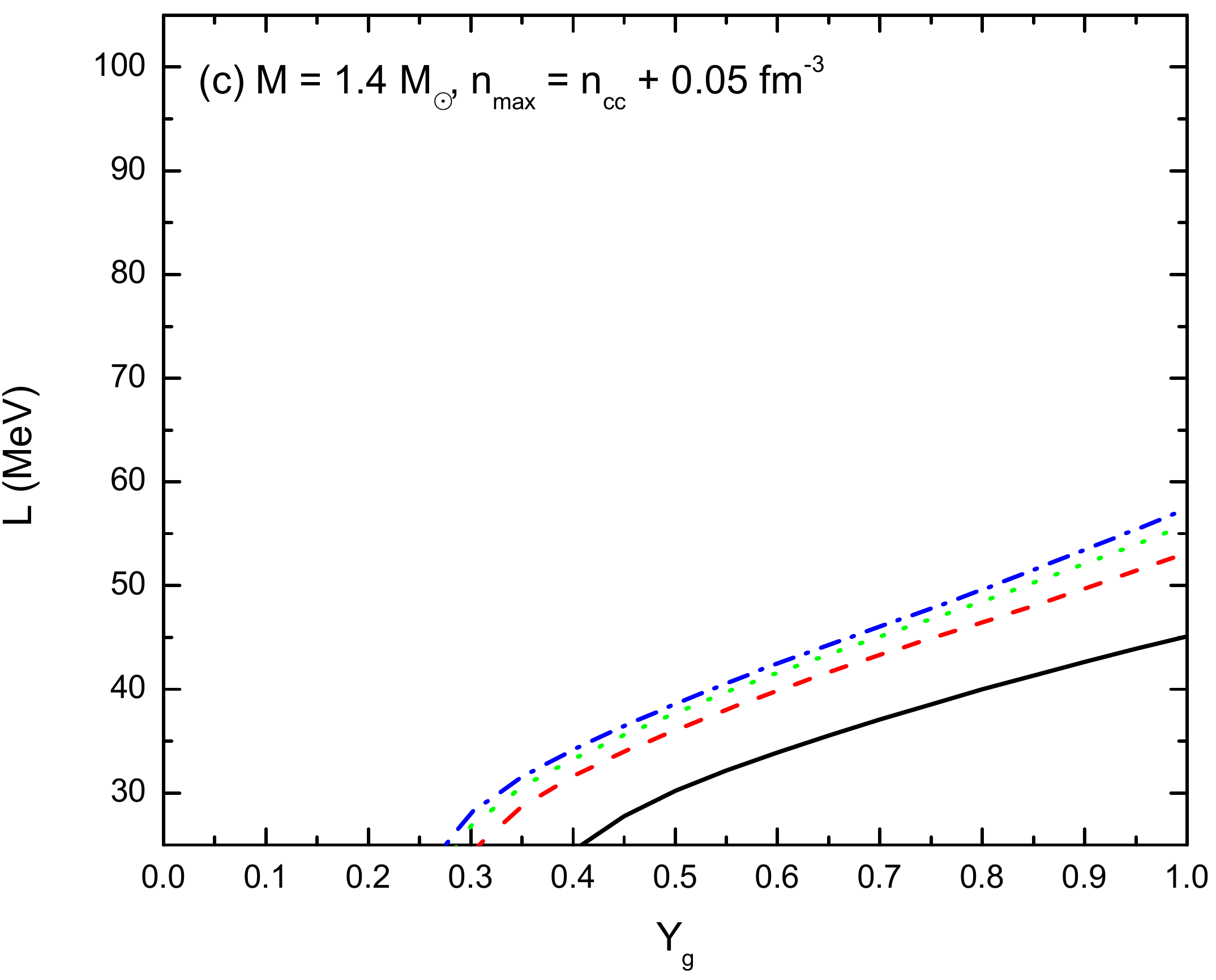}
\includegraphics[width=6cm,height=5cm]{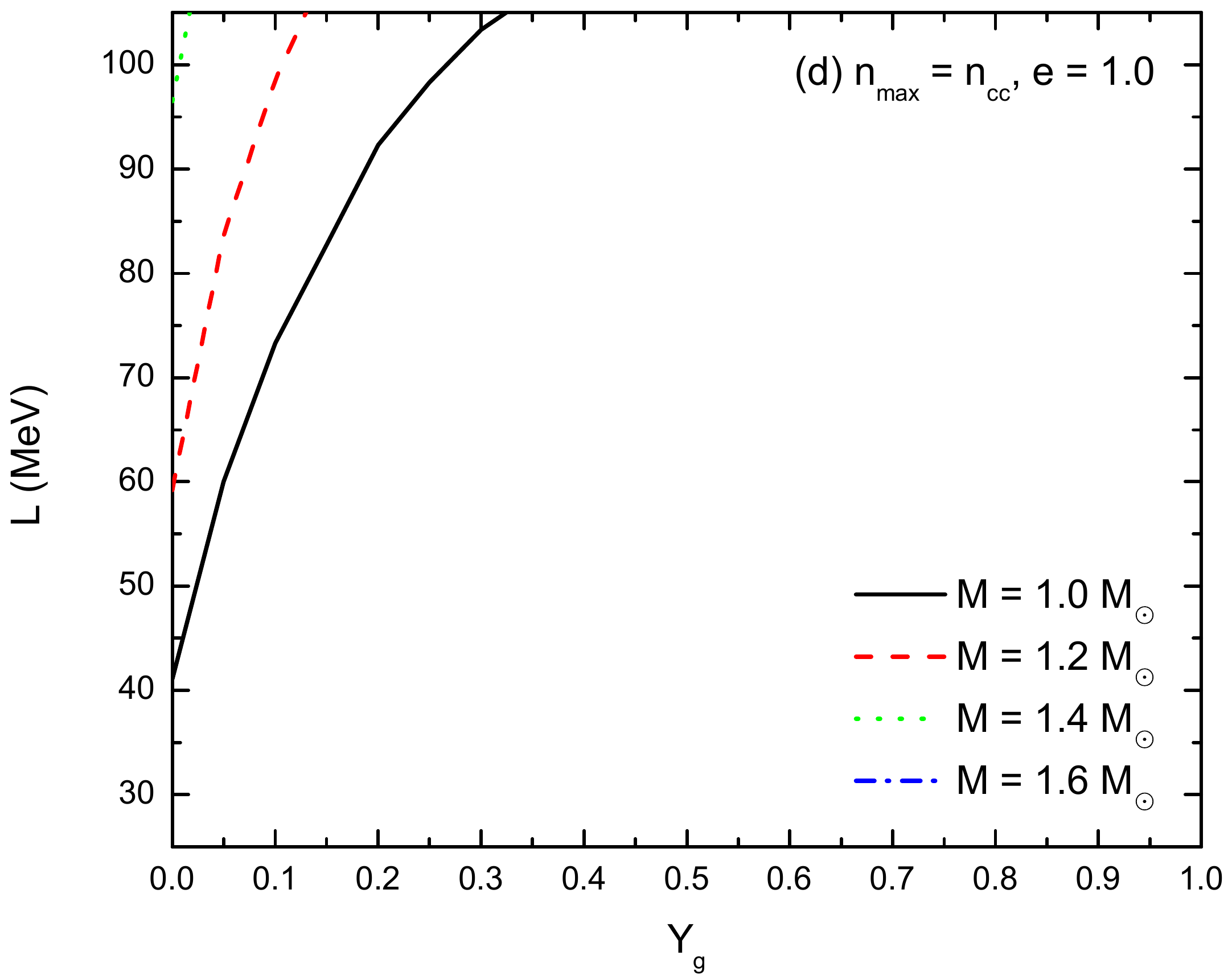}\includegraphics[width=6cm,height=5cm]{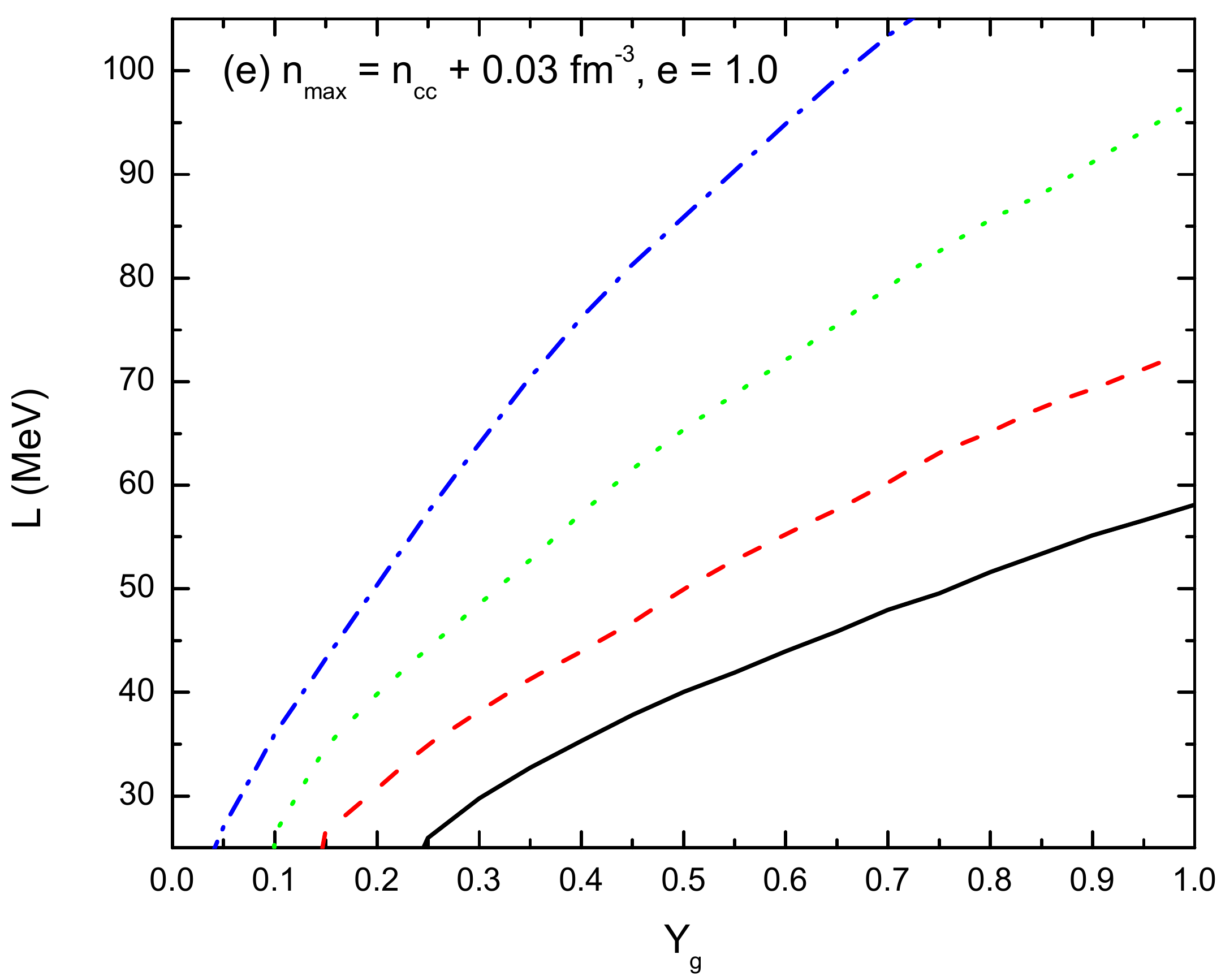}\includegraphics[width=6cm,height=5cm]{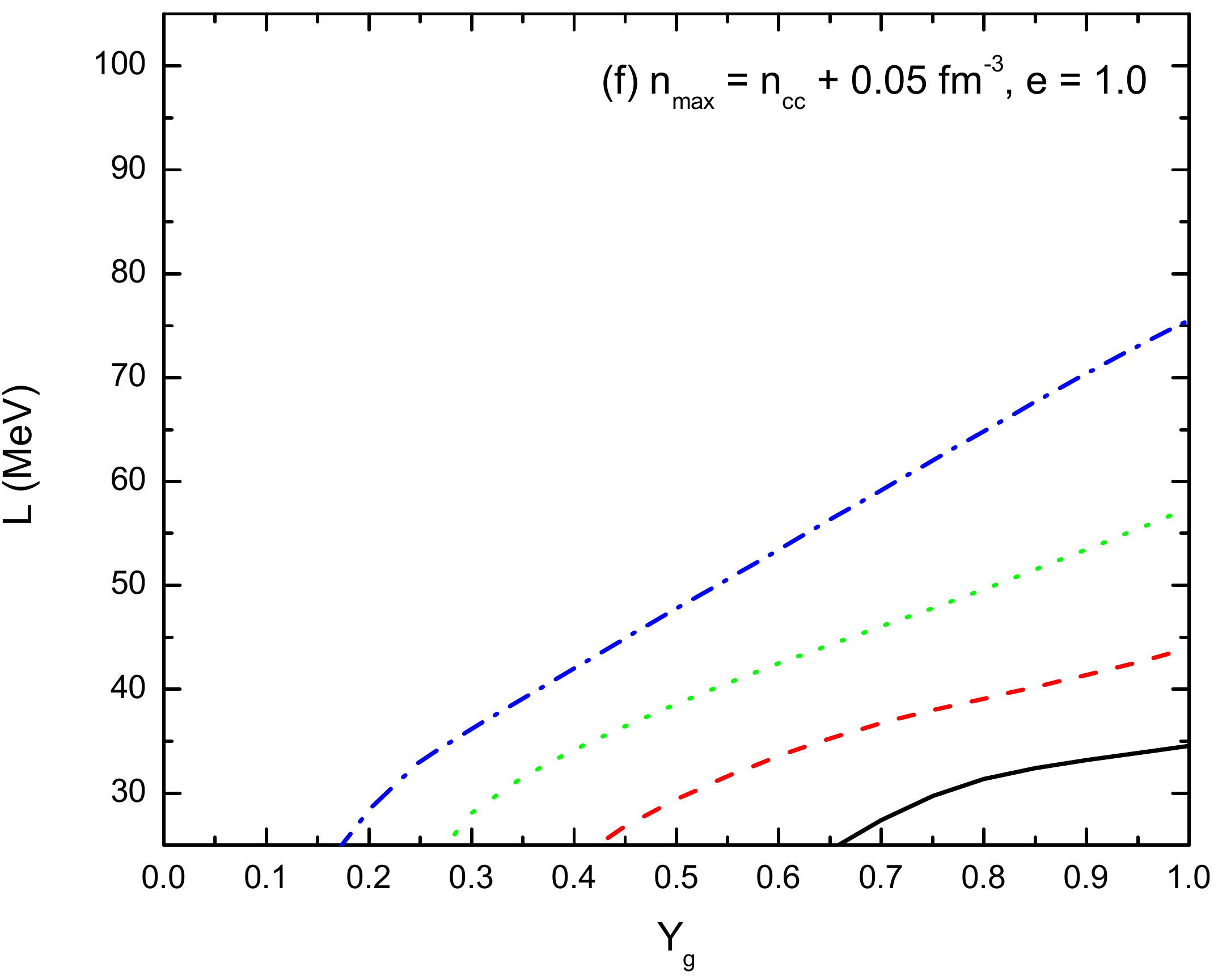}
\caption{(Color online) $Lower$ limits to the slope of the symmetry energy $L$ which satisfy the observational limit of $G\geq1.6\%$ for a given value of the core superfluid neutron fraction that is coupled to the crust at the time of glitch, $Y_{\rm g}$. Panels (a-c) show results for a 1.4$M_{\odot}$ star with crustal entrainment ignored $e=0$ (solid colored line), set to a strength $e$=0.33 (dashed line), 0.67 (dotted line) and 1 (dash-dotted line) and for a strong pinning region that is contained only within the inner crust $n_{\rm max}$=$n_{\rm cc}$ (a), and that extends into the outer core by a density of 0.03 fm$^{-3}$ (b) and 0.05 fm$^{-3}$ (c). Panels (d-f) show results for entrainment at full strength $e=1$, neutron star masses of $1.0$ (solid line), $1.2$ (dashed line), $1.4$ (dotted line) and $1.6M_{\odot}$ (dash-dotted line) and for a strong pinning region that is contained only within the inner crust $n_{\rm max}$=$n_{\rm cc}$ (d), and that extends into the outer core by a density of 0.03 fm$^{-3}$ (e) and 0.05 fm$^{-3}$ (f). Observationally consistent regions of parameter space lie to the \emph{left} of a given curve. Taken from \cite{Hooker2013}.}
\end{center}
\end{figure*}

%==============================================================================

What constraints do we have on $Y_{\rm g}$? In regions in which the protons form a Type I superconductor, neutron vortices can become magnetized as they entrain protons; then electron scattering off the magnetozed neutron vortices couples them to the charged component on timescales of $\sim 10 - 1000$s for the Vela pulsar \cite{Alpar1984g,Alpar1988g}. In regions where protons are Type II superconductors, they form fluxtubes which can pin neutron vortices, again coupling them to the charged component on timescales of days \cite{Babaev2009g,Link2012g}. By comparing the above estimates of coupling timescales with the upper limit of the glitch rise time $\lesssim 40$s, one may infer that only some fraction of the core neutron superfluid will contribute to the charged component of the star at the time of glitch.  Because that fraction is quite uncertain, we take $Y_{\rm g}$ to be a free parameter, but note that above estimates indicate that it is possible to have $Y_{\rm g} \ll 1$ \cite{Haskell2011g,Link2012g}. 

Let us also denote the total neutron fraction of the core at a given radius $r$ by $Q(r)$. Then the moment of inertia of the charged component can be expressed \cite{Seveso2012g}

\begin{align} \label{eq:MoI1}
I_{\rm c}=& \frac{8 \pi}{3} \int_0^R r^{4} [1 - Q(r) (1 - Y_{\rm gl})] \times \notag \\
& e^{-\nu(r)}\frac{\bar{\omega}(r)}{\Omega}\frac{\left( \mbox{\Large{$ \varepsilon $}} (r)+P(r) \right)}{\sqrt{1-2GM(r)/r}}\mathrm{d}r,
\end{align}

\noindent Denoting the energy density of crustal superfluid neutrons as $\mbox{\Large{$ \varepsilon $}}_{n} (r)$ and their pressure $P_{n}(r)$, and defining

\begin{equation} \label{eq:new2}
\mathcal{I} =\frac{2}{3} r^{2}e^{-\nu(r)}\frac{\bar{\omega}(r)}{\Omega}\frac{\left( \mbox{\Large{$ \varepsilon $}}_{n} (r)+P_{n}(r) \right)}{\sqrt{1-2GM(r)/r}}
\end{equation}

\noindent we can write the moment of inertia of the \emph{free} (non-entrained) crustal superfluid neutrons in the strong pinning (sp) region of the crust as

\begin{equation} \label{eq:new1}
I_{\rm csf}^{\rm (sp)}=\frac{2}{3} \int_{\rm sp \; region} {m_{\rm n} \over m_{\rm n}^*(r)} \mathcal{I} \mathrm{d} \mathcal{V}
\end{equation}

\noindent where $m_{\rm n}^*(r)$ is the mesoscopic effective neutron mass at radius $r$ in the crust, values of which are interpolated from the results of \cite{Chamel2012gb}. In order to examine systematically the effect of reducing the strength of entrainment in the crust, we introduced the parameter $e$ such that $e$=0 represents  no entrainment and $e$=1 represents full strength entrainment: $m_{\rm n}^* \to 1 + (m_{\rm n}^* - 1)e$. We can then calculate the ratio $I_{\rm csf}^{\rm (sp)}/I$ as a function of the stellar mass $M$, the slope of the symmetry energy $L$, the fraction of core neutrons coupled to the crust during the glitch $Y_{\rm g}$ and the strength of entrainment in the crust $e$, and confront the result with the observational requirement that 

\be
I_{\rm csf}^{\rm (sp)}/I \geq 0.016.
\ee

For each value of $Y_{\rm g}$ one can identify the value of $L$ for which $G\geq1.6\%$ is satisfied below a given mass of neutron star. Fig.~7  plots those values of $L$ vs $Y_{\rm g}$; only values of $Y_{\rm g}$ \emph{to the left} of the curves satisfy the observational constraint. In Fig.~7 (a-c), we show the values of $L$ and $Y_{\rm g}$ that satisfy $G\geq1.6\%$ for stars with a mass of $1.4 M_{\odot}$ and different levels of entrainment $e=$0, 0.33, 0.67 and 1.0. Without crustal entrainment $e=0$, fractions of core superfluid coupled to the crust on glitch rise timescales up to $Y_{\rm g}$=0.75 for $L=105$ MeV satisfy $G\geq1.6\%$ for a 1.4$M_{\odot}$ star. Including entrainment at the level of $e=0.33$ limits $Y_{\rm g}<$0.2; at $e=0.67$, $Y_{\rm g}<$0.07, and at $e=1.0$, $Y_{\rm g}<$0.02. In the latter case of full entrainment, consistency with observation is reached only for very high values for $L \geq$100 MeV. Because of this, we also tested the possibility that some of the core neutrons are also involved in the angular momentum transfer by extending the strong pinning region into the core by densities of 0.03 fm$^{-3}$ Fig.~7(b), and 0.05 fm$^{-3}$ Fig.~7(c). Extending pinning into the core by at least 0.03fm$^{-3}$ above the crust-core transition density allows us to find values of $L$ for which $G\geq1.6\%$ is satisfied for any value of $Y_{\rm g}$ ($L>95$ MeV for $Y_{\rm g}=1.0$); extending pinning by 0.05fm$^{-3}$ allows us to satisfy $G\geq1.6\%$ at $e=1.0$ for $L>57$ MeV for any $Y_{\rm g}$. In Figs.~7(d)-(f) the mass dependence of the results is shown taking into account entrainment at full strength. Fig.~7(d) shows that, of Vela was a low mass neutron star $\leq 1.2 M_{\odot}$, the observational constraints would be met for $L>60$ MeV for $Y_{\rm g} \ll 0$ and full entrainment.

The main results of this analysis can be summarized as follows. The measured value of $G$ from the last 4 decades worth of pulse timing data from the Vela pulsar generally favors a larger value for the symmetry energy slope $L$ (stiffer symmetry energy) because, for this set of EoSs, a larger $L$ tends to produce neutron stars with larger crusts relative to their cores, and hence larger $\Delta I/I$ if $\Delta I$ is associated with a component of the crust. We specifically identify the components $\Delta I$ and $I$ broadly with the components involved in the ``snowplough'' glitch model, $I_{\rm csf}^{\rm (sp)}$ and $I_{\rm c}$ respectively. Neglecting the entrainment of crustal superfluid neutrons by the crustal lattice, the constraint on $G$ is found to be satisfied for every value of $L$ between 25 and 105 MeV, with an upper limit to the fraction of core neutrons coupled to the crust during the glitch event which increases from $\approx 5\%$ at $L=$25 MeV up to $\approx 65\%$ at $L=105$ MeV. However, with crustal entrainment at the level predicted by microscopic calculations, the constraint on $G$ is satisfied only with $L>$100 MeV and $Y_{\rm g} \ll 1$. This analysis, combined with previous studies \cite{Chamel2012gb,Andersson2012g}, indicate that, taking into account the hydrodynamic interaction between the crustal superfluid and lattice, the crustal glitch paradigm is incompatible with observation unless the symmetry energy at saturation density, and hence the EoS, is very stiff. One way to resolve this situation is if the region of strong pinning is allowed to penetrate some way into the core; we show that if it penetrates to densities up to twice the crust-core transition, then consistency with observations is attained for values of $L$ down to 25 MeV. A detailed physical account of this scenario is yet forthcoming.\\

\noindent \textbf{\emph{Constraint:}} $L\gtrsim 100$ MeV.

\noindent \textbf{\emph{Caveats:}} Crust-initiated glitch paradigm still incomplete (notably the trigger mechanism for the glitch is quite uncertain). Core components could be involved in the glitch process. Further progress in falsifying models likely to involve detailed comparison with post-glitch spin evolution of pulsars.

%%%%%%%%%%%%%%%%%%%%%%%%%%%%%%
%
% SGR QPOs
%
%%%%%%%%%%%%%%%%%%%%%%%%%%%%%%

\section{Quasi-periodic oscillations in the light curves of giant flares from Soft Gamma-ray Repeaters}
\label{sec:5}

One observational class of neutron star, soft gamma-ray repeaters (SGRs) exhibit regular gamma-ray flares with total energies of $\sim 10^{41}$ ergs, durations of around 0.1s, and a pattern of occurrence statistically consistent with self-organized critical phenomena such as Earthquakes \cite{Cheng1996} which is suggestive of an origin in crustal magnetic and seismic activity \cite{Thompson1995,Thompson1996,Kondratyev2002}. Occasionally they exhibit giant flares with total energy outputs of $10^{44} - 10^{46}$ ergs. One such flare led to the discovery of SGRs in 1979 \cite{Mazets1979} from what is now designated SGR 0525-66; a marginal detection of a 43Hz quasi-periodic oscillation (QPO) oscillation in its aftermath was reported in 1983 \cite{Barat1983}.  A 2004 giant flare from SGR 1806-20 \cite{Hurley2005} led to a significant detection of several QPOs in the X-ray tail of the flare's light curve \cite{Israel2005,Watts2006,Strohmayer2006,Hambaryan2011} at frequencies of 18, 26, 30, 93, 150, 626 and 1837Hz. Subsequently, analysis of data taken from a third giant flare, from SGR 1900+14, which occurred in 1998 \cite{Hurley1999}, revealed the presence QPOs in its light curve \cite{Strohmayer2005} at frequencies of 28, 53, 84 and 155Hz.

SGRs are believed to be one observational realization of magnetars, neutron stars with very high magnetic fields, up to $\sim 10^{15}$G, and it is the decay of these high fields that are believed to power giant flares \cite{Thompson1995,Thompson1996}. The standard paradigm of giant flares model posits that the release of magnetic energy in giant flares is triggered by crust fracturing after the crust is stressed to breaking point by the magnetic field anchored to it \cite{Thompson1995,Thompson1996}. The subsequent quasi-periodic modulation of the post-flare light curve is interpreted as being, at least in part, a result of torsional crust oscillations \cite{Duncan1998,Messios2001,Piro2005,Bastrukov2007,Samuelsson2007}. This offers a tantalizing probe of crust properties such as thickness and global neutron star properties such as mass and radius, and hence the nuclear symmetry energy, through neutron star seismology \cite{Samuelsson2007,Steiner2009}. 

It has been demonstrated, for example, that varying mass $M$ and radius $R$ of a stellar model (with fixed crust model) to match a particular fundamental mode frequency produces an orthogonal correlation between $M$, $R$, than matching the first overtone does. Hence, simultaneous matching of fundamental and first overtone has the potential to tightly constrain $M$ and $R$ \cite{Samuelsson2007,Lattimer2007,Deibel2013}, and hence the EOS.

Torsional modes are designated $_{\rm n}t_{\rm l}$ where $n$ and $l$ are the number of radial and angular nodes associated with the mode. Conservation of angular momentum dictates that $l \ge 2$. We can explicitly demonstrate the dependence of the mode frequencies on crustal and stellar properties using a simple analytic approximation assuming and isotropic crust \cite{Samuelsson2007}. The frequency of the fundamental modes $_{\rm 0}t_{\rm l}$, $\omega_0$, and of the overtones $_{\rm n \neq 0}t_{\rm l}$,  $\omega_n$, can be written

\begin{align}
	&\displaystyle\omega^2_0 \approx \frac{e^{2\nu}v_s^2(l-1)(l+2)}{2RR_c}, \notag \\
	& \omega^2_n \approx e^{\nu - \lambda} {n \pi v_s \over \Delta} \bigg[ 1 + e^{2\lambda} {(l-1)(l+2)\over 2 \pi^2} {\Delta^2 \over R R_c} {1 \over n^2},\bigg]
\end{align}

\noindent where $n,l$ are the number of radial and angular nodes the mode has respectively. $M, R, R_{\rm c}$ and $\Delta$ are the stellar mass and radius, the radius out to the crust-core boundary and the thickness of the crust respectively. The symmetry energy dependence clearly enters through the strong correlation of $R$ (and hence $R_{\rm c}$) and $\Delta$ with the slope of the symmetry energy $L$.

$v_{\rm s}$ is the shear speed at the base of the neutron star crust and $\nu$ and $\lambda$ are metric fields. $v_{\rm s}^2 = \mu/\rho$ where $\rho$ is the mass density and $\mu$ is the shear modulus at the base of the crust, which, assuming the crust has the structure of a Coulomb crystal, has a form extracted from molecular dynamics simulations of \cite{Ogata1990,Strohmayer1991,Hughto2008}

\be \label{eq:shear_mod}
	\mu_{\rm s} = 0.1106 \left(\frac{4\pi}{3}\right)^{1/3} A^{-4/3} n_{\rm b}^{4/3} (1-X_{\rm n})^{4/3} (Ze)^2.
\ee

%====================================== FIGURE 8 ===============================

\begin{figure}
\begin{center}
\includegraphics[width=20pc]{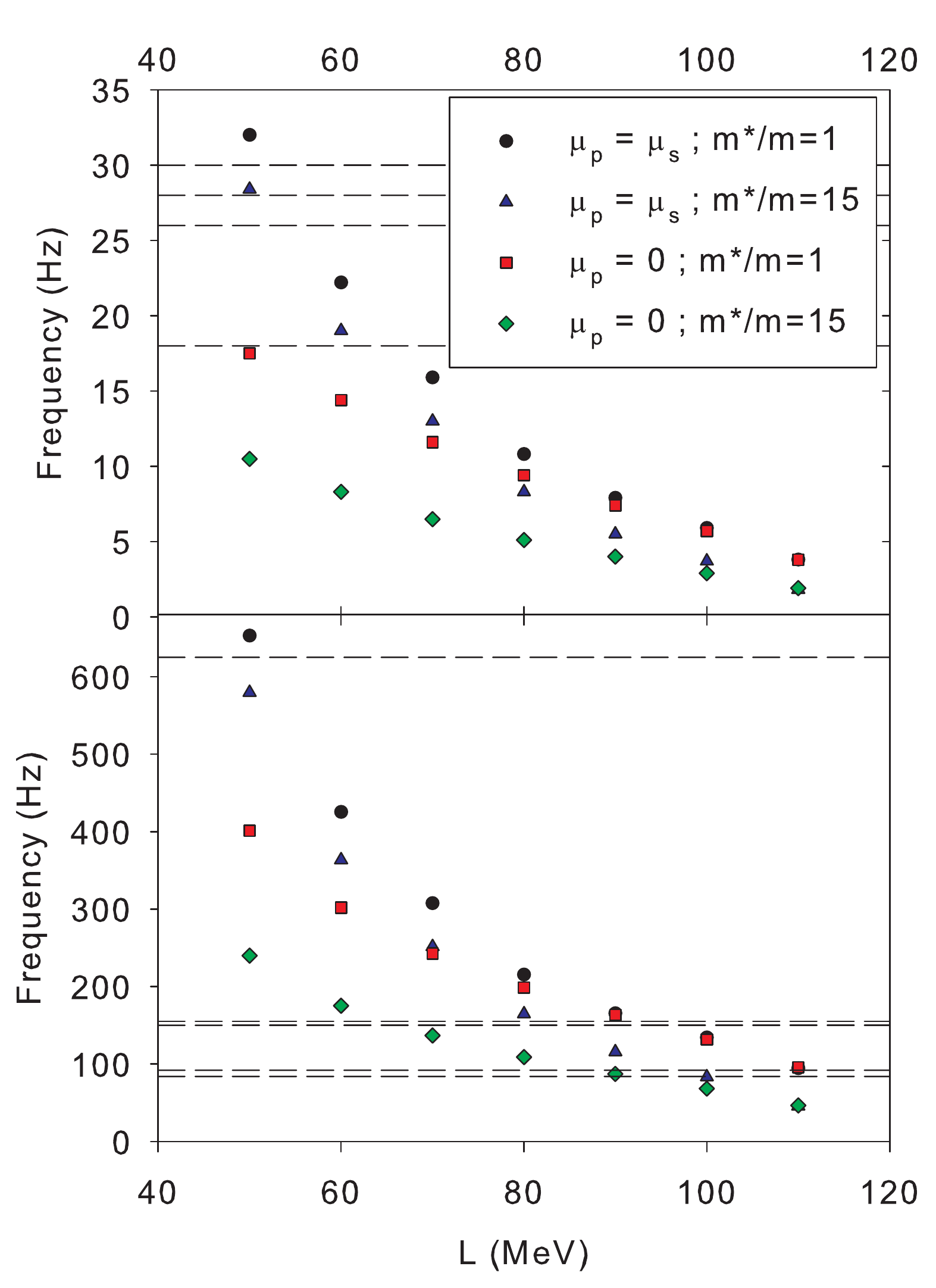}
\caption{\label{fig:8} (Color online) Fundamental ($_0t_2$, top) and first overtone ($_1t_2$, bottom) frequencies of torsional crust oscillations as a function of the slope of the symmetry energy at saturation density, $L$, for our family of EOSs fit to \emph{ab-initio} pure neutron matter calculations. Circles and triangles represent results with the shear modulus of pasta set to that of an elastic solid, while squares and diamonds represent results with the shear modulus set to zero. Circles and squares represent the cases where entrainment of crustal superfluid neutrons is neglected, while triangles and diamonds represent the case where it is taken into account. The observed QPOs from SGR flares that fall in the displayed frequency ranges are shown by the horizontal dashed lines. Taken from \cite{Gearheart2011}.}
\end{center}\hspace{2pc}%
\end{figure}

%==============================================================================

\noindent Here, $A$ and $Z$ are the mass and charge numbers of the nuclei at the base of the crust (which occurs at a baryon density $n_{\rm b}$) and $X_{\rm n}$ is the density fraction of free neutrons there. Through the shear modulus, the torsional mode frequency spectrum also depends on the microscopic structure and composition of the crust which is also dependent on the symmetry energy \cite{Newton2013}.

A first study of the link between symmetry energy related nuclear observables and torsional crust modes using the above forms of the frequencies concluded that, to identify the observed frequencies at around 30Hz with the fundamental torsional mode $_{\rm 0}t_{\rm 2}$ requires a soft symmetry energy (small $L$ at saturation density) \cite{Steiner2009}. We further quantified this using our consistent crust-core EOSs and the same forms for the frequencies of the torsional modes \cite{Gearheart2011} to examine explicitly their dependence on the slope of the symmetry energy $L$. We also examined the limiting effects of the existence of the pasta phases by treating the shear modulus of the pasta phases in the limits of (i) the pasta having elastic properties similar to the rest of the solid crust, therefore allowing the use of the shear modulus $\mu_{\rm s}$ as given above at the crust-core transition density (a case we refer to as ``solid pasta''), and (ii) the pasta behaves as a fluid with no shear viscosity, which, in this model, effectively lowers the density at which we evaluate the shear modulus and crustal frequencies from the crust-core transition density to the density at which spherical nuclei make the transition to the pasta phases (a case we refer to as ``fluid pasta''). We compared the maximal effect of pasta to the effect of superfluid neutrons scattering off the crustal lattice by multiplying the frequencies by a mesoscopic effective mass term $\epsilon_{\star} = { (1 - X_{\rm n}) /[1 - X_{\rm n} (m_{\rm n} / m_{\rm n}^*)]}$ where $m_{\rm n}^*$ is the mesoscopic effective neutron mass \cite{Andersson2009sgr}. Physically, the crustal lattice is having to drag along a certain fraction of the dripped neutrons as it oscillates (in fluid mechanics parlance, the lattice entrains the superfluid neutrons), and the frequency is reduced.  We took as extreme values $m_{\rm n}^* / m_{\rm n}$ = 1 (no entrainment) and $m_{\rm n}^* / m_{\rm n}$ = 15 (maximum entrainment) \cite{Chamel2005g}. 

We show the results we obtained in \cite{Gearheart2011} for the dependence on $L$ of the fundamental (upper panel) and first overtone (lower panel) frequency in Fig.~8. Generically, the frequency decreases as $L$ increases because the radius $R$, in the denominator of Eq.~13, increases with $L$. Circles and squares give the frequencies neglecting entrainment for the solid pasta and fluid pasta respectively, illustrating that the effect of pasta is to reduce the frequencies by a factor of $\approx 2$. The triangles and diamonds refer to the same two cases, but with entrainment taken into account. Inclusion of entrainment reduces the frequencies by about 20-30\% at most, and so is a somewhat smaller effect than that of the pasta phases. The horizontally dashed lines indicate the measured frequencies from SGRs. Allowing for uncertainties in the model, which will be discussed a little later, we demand only that our calculated fundamental frequencies fall somewhere in the range of 18 - 30Hz, the set of candidate fundamental frequencies observed. This is obtained only if $L \lesssim 60$ MeV. If, additionally, we demand that the observed 625Hz mode is to be matched to the 1st overtone, we then require that $L \lesssim 60$ MeV \emph{and} the pasta phases should have mechanical properties approaching that of an elastic solid. 

One might attempt an exact identification of all observed frequencies with theoretically calculated modes, employing the full spectrum of fundamental and overtones. A series of papers have attempted just this \cite{Sotani2012,Sotani2013a,Sotani2013b}. A similar formalism for calculating the torsional modes is used, and as in our study, a set of crust compositions and EOSs derived as a function of $L$  are employed. However, the core radius and mass is fixed in these studies, preventing the additional correlations of $L$ with bulk stellar properties to be folded into the resulting relationship between $L$ and the mode frequencies. In \cite{Sotani2012}, the entrainment effect is neglected and the shear modulus of pasta is set to zero, as in our ``fluid pasta'' case. A similar dependence of $L$ on the frequencies of the fundamental and first overtone modes is found, and by taking the lowest observed frequency, 18Hz from SGR 1806-20, to be the fundamental, they obtain $L \gtrsim 50$ MeV. In \cite{Sotani2013a}, using the same model but with superfluid entrainment effects are included and the full range of observed frequencies from both SGR1806-20 and 1900+14 is simultaneously matched to various fundamental frequencies, obtaining then a constraint $100\lesssim L \lesssim 130$ MeV, notably at odds with most experimental constraints on $L$. Mindful of this, \cite{Sotani2013b} note that by excluding the second-lowest frequency observed in SGR 1806-20 from the analysis, one then obtains $58 < L < 85$ MeV.
 
The models used in obtaining the above constraints neglect the effect of the stellar magnetic field, which, considering we believe we are dealing with magnetars, should be cause for concern. There are three main areas where the magnetic field can play an important role in the interpretation of the frequencies. Firstly, the lowest observed frequencies can also be explained as fundamental modes or first overtones of the Alfven spectrum \cite{Levin2008,Sotani2008,Colaiuda2009,CerdaDuran2009} (and then can potentially place constraints on the magnetic field geometry \cite{Sotani2008b,Gabler2013a}). Secondly, the crust modes couple to the Alfven modes, potentially modifying the crust modes \cite{Lee2007,Sotani2007,vanHoven2011} or rapidly damping them out \cite{Gabler2011,Gabler2012,Gabler2013a,Gabler2013b}. Thirdly, the magnetic field can affect the composition of the crust itself \cite{Nandi2013}. The coupling of the crust to core means that the exact superfluid and superconducting states of matter there are highly relevant, and lead to a further enriching of the possible mode spectrum and decay modes \cite{Gabler2013b,Glampedakis2013}. Nevertheless, the imprint of the stellar crust and the radius and moment of inertia of the star on the mode spectrum will still be retained, and their dependence on the symmetry energy and its slope $L$ cannot be ignored if we want future precision asteroseismology to come to fruition. \\

\noindent \textbf{\emph{Constraints (caveats):}} (i) $L\lesssim 60$ MeV (Theoretical frequencies fall in range of observed QPO frequencies; consistent crust \& core EOSs; limiting pasta and superfluid effects crudely taken into account.) (ii) $L \gtrsim 50$ MeV (Exact matching of lowest observed frequency to fundamental mode; inconsistent modeling of crust, core; fluid pasta; no superfluid effects.) (iii) $100\lesssim L \lesssim 130$ MeV (Exact matching of all observed frequencies; inconsistent modeling of crust, core; fluid pasta; superfluid effects included.) (iv) $58 < L < 85$ MeV (Exact matching of all observed frequencies except second lowest in SGR1806-20; inconsistent modeling of crust, core; fluid pasta; superfluid effects included.)

\noindent \textbf{\emph{General caveats:}}  Coupling to Alfven modes ignored. Low frequency modes can be explained by pure Alfven modes.

%%%%%%%%%%%%%%%%%%%%%%%%%%%%%%
%
% ACCRETION INDUCED SPIN UP AND R-MODES
%
%%%%%%%%%%%%%%%%%%%%%%%%%%%%%%

\section{Lower limit to the observed spin periods of old, recycled pulsars}
\label{sec:6}

The population of old neutron stars, evolved (or still evolving) through accretion-induced spin-up caused by interaction with a binary companion, is observed to rotate with periods up to milliseconds.The maximum rotation frequency of any known pulsar is 716 Hz \cite{Hessels2006}. Rapidly rotating neutron star models constructed with reasonable EoSs tend to give theoretical maximum spin rates of $\sim 2000$ Hz and above, beyond which material will be ejected from the equator of the star. This suggests that neutron stars are not continually spun-up by accretion torques until they reach their break-up frequency, but rather a counter-torque is encountered at substantially frequencies that prevents any further spin-up \cite{Chakrabarty2003,Patruno2012}.

An origin for the counter-torque that has been intensively studied over the past decade and a half involves gravitational radiation reaction from a class of inertial oscillation modes called r-modes that might be driven unstable. r-modes are analogous to Rossby waves in Earth's atmospheres and oceans as they have the Coriolis force as their restoring force. Below a certain spin period, r-modes can become unstable through the Chandrasekhar-Friedman-Schutz (CFS) mechanism \cite{Chandrasekhar1970,Friedman1978} in which gravitational radiation from the modes drives the oscillations to higher amplitudes. The mode becomes unstable if the timescale for the CFS mechanism to drive the r-modes unstable is shorter than the timescale for the modes to be damped by the internal viscosity of the star; in order to explain the observed cutoff, this should occur at frequencies just a little higher than the highest observed frequency  \cite{Bildsten1998,Friedman1998,Andersson1999}. There are various sources of viscosity that are, in general, temperature dependent; for a given temperature, equating the gravitational radiation and viscosity timescales $\tau_{\rm GR} = \tau_{\rm v}$ gives the frequency above which stars become unstable. In frequency-temperature space, therefore, there is a division between a region where neutron stars are stable and a region in which they are not (the instability window). Estimates for the internal temperature of millisecond pulsars can be obtained from X-ray and Ultra-violet observations assuming accretion heating is balanced by core neutrino emission. If millisecond pulsars for which estimates of their internal temperature exist are observed to fall within the instability window, then our physical picture is incomplete.

The r-mode amplitude grows towards the surface of the star, and hence the crust is expected to play an important role in determining the stability of stars. This is where the role of the symmetry energy in determining crustal properties comes into play, and two studies have utilized this to set constraints on $L$. Both studies use a simple model in which the crest is assumed to be perfectly rigid, and at the temperatures estimated for the interiors of millisecond pulsars $\sim 1-5 \times 10^8$K, the dominant source of viscosity is the shear viscosity resulting from electron-electron scattering.

Let us briefly discuss the results of our study. The timescales $\tau_{\rm GR}$, $\tau_{\rm v}$ (the latter in the case of shear viscosity from e-e scattering) are given by \cite{Lindblom1998,Lindblom2000,Owen1998}.

\begin{equation}\label{TGR}
 {1\over\tau_{GR}} =  {32\pi G \Omega^{2l+2}\over c^{2l+3}}
{(l-1)^{2l}\over [(2l+1)!!]^2}  \left({l+2\over
l+1}\right)^{2l+2} \int_0^{R_c}\rho r^{2l+2} dr,
\end{equation}

\begin{equation}\label{TV}
 \tau_v  = \frac{1}{2\Omega} \frac{{2^{l+3/2}(l+1)!}}{l(2l+1)!!{\cal
 I}_l}
\sqrt{2\Omega
R_c^2\rho_c\over\eta_c} \int_0^{R_c}
{\rho\over\rho_c}\left({r\over R_c}\right)^{2l+2} {dr\over R_c},
\end{equation}

\noindent  Here, the viscosity is evaluated at the crust-core boundary where the shear damping rate is greatest \cite{Bildsten2000,Andersson2000a,Lindblom2000,Rieutord2001}. $\rho_{\rm c}$ is the crust-core transition density and $R_{\rm c}$ the stellar radius at that density and $\nu_{\rm c}$ the viscosity there; it is here that the symmetry energy dependence on the crustal thickness enters the analysis. We consider the case of $l=2$, with $\mathcal{I}_2$ = 0.80411. The shear viscosity resulting from the electron-electron scattering process $ \eta_{ee}=6.0\times 10^6 \rho^2 T^{-2}$~(g cm$^{-1}$ s$^{-1})$ \cite{Flowers1979,Cutler1987}. By employing our consistent set of crust and core EoSs, we can calculate the frequency of onset of the CFS instability as a function of temperature for different values of $L$: the results of such calculations are plotted in Fig.~9 for a 1.4$M_{\odot}$ and 2.0$M_{\odot}$ star (upper and lower panels respectively) \cite{Wen2012}. We will focus only on the high mass neutron star results since ms pulsars have been accreting material for a large fraction of their lives.

%====================================== FIGURE 9 ===============================

\begin{figure}[t]
\begin{center}
\includegraphics[width=18pc]{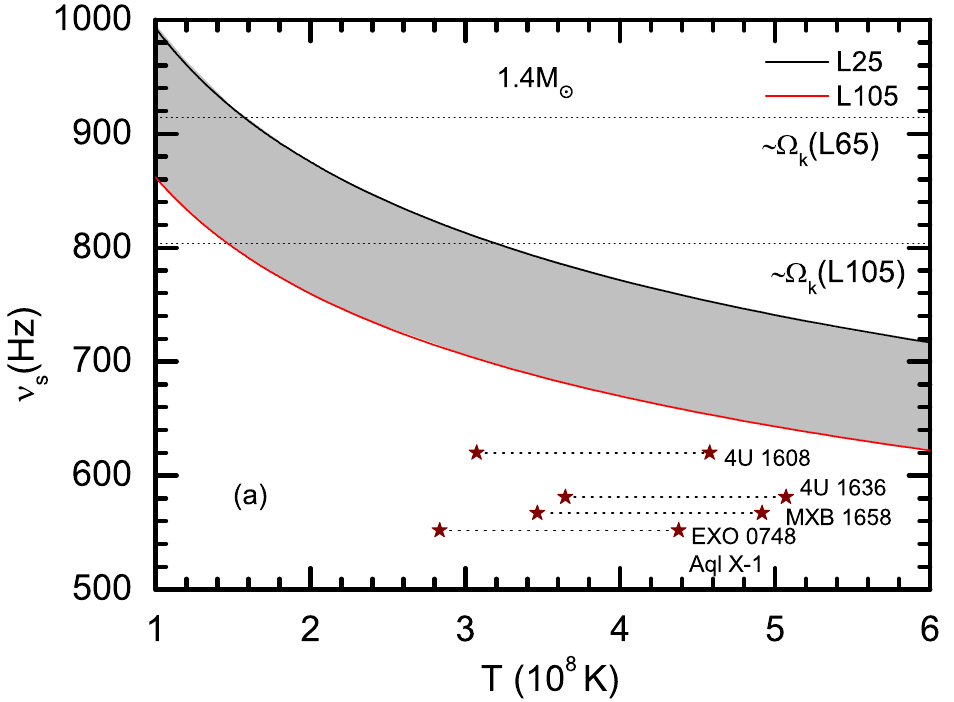}
\includegraphics[width=18pc]{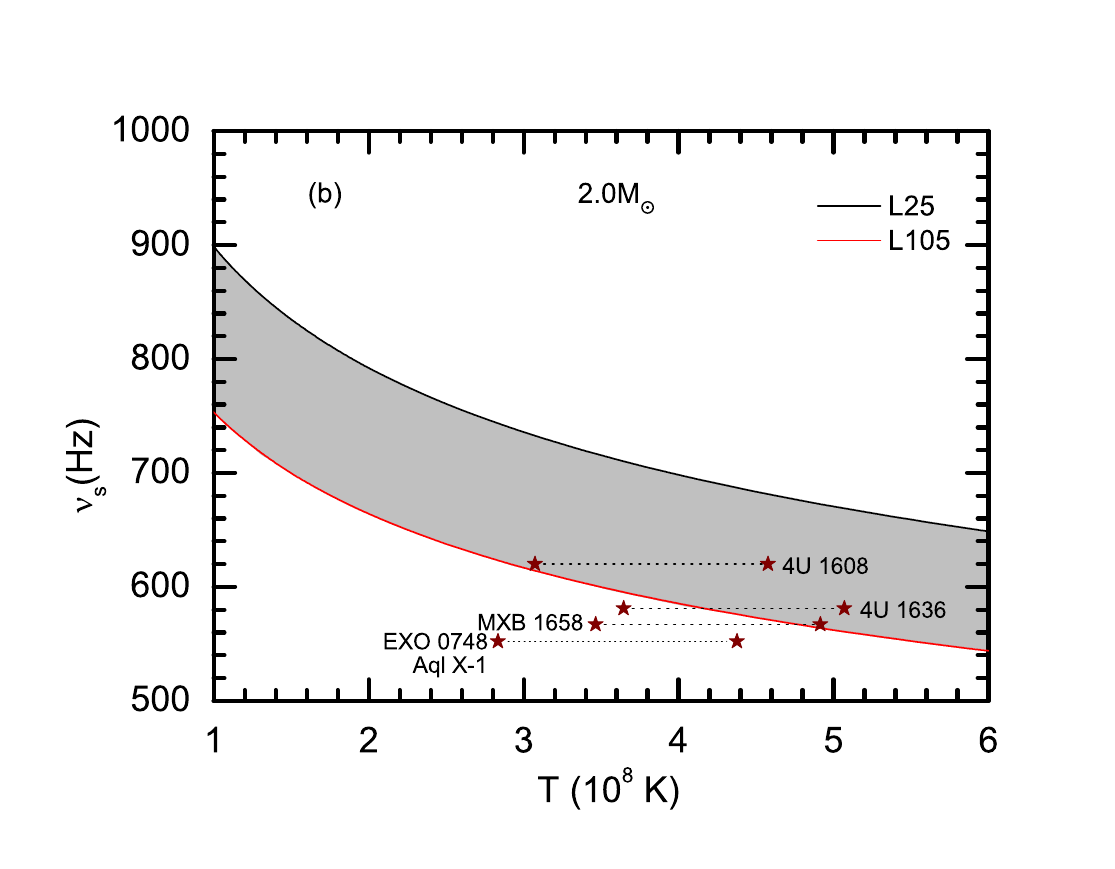}
\caption{\label{fig:9}Frequency above which the electron-electron viscosity is insufficient to damp the gravitational radiation-driven r-mode instability versus core neutron star temperature for a 1.4$M_{\odot}$ star (left) and 2.0$M_{\odot}$ (right). Results are shown for the two bounding EoSs in our PNM sequence: $L=25$MeV and $L=115$MeV, with the region in between shaded in grey. The positions of 4 millisecond pulsars are shown with two estimates of their internal temperature; one interpretation of observations is that the frequency of onset of the r-mode instability should be higher than any observed pulsar frequency at a given temperature in order to be consistent with those observations. For comparison, the neutron star break-up frequency for $L=25, 65$ and $105$ MeV are shown if they fall within the frequency range of the plot. Taken from \cite{Wen2012}.}
\end{center}
\end{figure}

%=============================================================================

The solid curve shows the frequency above which the r-mode becomes unstable for $L$=25 MeV; the dotted line for $L$=115 MeV. The location of four neutron stars in short recurrence time LMXBs are shown by the star symbols; the temperature has been estimated in two different ways \cite{Watts2008,Keek2010} leading to upper and lower bounds. Let us focus on the object 4U 1608. For the range of estimated temperature, the object is clearly in the instability window for $L=105$ MeV. It should therefore spin down rapidly, and we should not observe it to have as high a frequency as we do. Therefore, to reconcile the model with observations, we must exclude the $L=105$ MeV EoS. By this argument, and adopting a conservative requirement that the estimated temperature interval for 4U 1608 fall completely below the instability window, we must require that $L<65$ MeV.

A similar investigation, adopting the rigid crust model, concludes that $L>50$ MeV \cite{Vidana2012}. There are two important differences between this study and ours: (i) viscous damping is taken to act throughout the whole of the interior rather than in a viscous boundary layer at between the crust and the core, and (ii) the neutron star core and crust are not consistently modeled: the radius of the star is varied independently of $L$, when in fact there is a strong correlation between the two. Nevertheless, both studies demonstrate that the position of the r-mode instability window is quite sensitive to the slope of the symmetry energy $L$.

Let us finally outline some of the multitude of uncertainties that have been evinced by r-mode modeling and by careful observational analysis of the millisecond pulsar systems themselves. Firstly, it is possible that the origin of the limiting spin period of pulsars has a physical origin not in the r-mode instability but by interactions between the pulsar and its accretion disc \cite{White1997,Andersson2005a,Patruno2012}. Secondly, the assumption of a rigid crust is not justified; it merely sets an upper limit on the viscous damping rate (and hence an upper limit to the frequency at which the r-mode goes unstable). However, treating the crust realistically (with a finite shear modulus), it is found that the r-modes can penetrate the crust and therefore the relative motion of the crust and the core fluid (known as the ``slippage'') is reduced, reducing significantly the viscous dissipation there \cite{Levin2001,Glampedakis2006}. This significantly reduces the maximum frequency a neutron star can be spun up to; the magnitude of the slippage and its frequency dependence are still sensitive to the crust thickness and hence the symmetry energy. Additionally, we would expect such as scenario to be sensitive to the existence of nuclear pasta. When modeled with the inclusion of crust-core slippage, or with other sources of viscosity such as mutual friction between superfluid and superconducting components of the core and hyperon bulk viscosity, most observed systems fall within the instability window \cite{Ho2011,Haskell2012r}, which indicates that our understanding of the r-mode instability and its dependent physics is incomplete; the sensitivity of the latter results to the crust thickness has yet to be investigated, however. One way out of the predicament would be that unstable r-modes saturate at amplitudes too small to be effectively damped \cite{Bondarescu2007,Haskell2013r}. \\

\noindent \textbf{\emph{Constraints (caveats):}} (i) $L\lesssim 65$ MeV (consistent crust, core EOSs; r-mode damping via shear viscosity in a viscous boundary layer at the crust-core transition) (ii) $L > 50$ MeV  (inconsistent crust, core EOSs; r-mode damping via shear viscosity throughout the core)

\noindent \textbf{\emph{General caveats:}}  Perfectly rigid crust unrealistic; r-modes likely penetrate the crust, reducing the crust-core slippage and hence the viscous dissipation there. r-modes instability might saturate at sufficiently low amplitudes to allow stars to inhabit the instability window. Other likely sources of viscosity from superfluid and exotic components ignored. Limit to frequency to which neutron stars can be spun-up to might have physical origin elsewhere, for example in star-accretion disc interactions.

%%%%%%%%%%%%%%%%%%%%%%%%%%%%%%
%
% BINARY NS MERGERS
%
%%%%%%%%%%%%%%%%%%%%%%%%%%%%%%

\section{Electromagnetic precursors to short gamma-ray bursts}
\label{sec:7}

One endpoint of the evolution of a high-mass binary system is a binary neutron star. In a certain fraction of such systems, the first-born neutron star (the primary) undergoes a common envelope (CE) evolution when the secondary star's envelope overspills its Roche lobe \cite{Bhattacharya1991}. During this phase, friction shrinks the separation of the stars significantly. The CE phase ends with the ejection of envelope. If the subsequent secondary supernova leaves behind a neutron star, the result will be a binary neutron star system whose separation is small enough that the stars will merge due to gravitational wave-induced orbital decay within the Hubble time. Electromagnetic (EM) and gravitational wave (GW) signals from neutron star mergers can provide a wealth of important astrophysical information including the nuclear equation of state (e.g., \cite{Bauswein2012}). Large uncertainties about the rate of production, and hence merger, of binary neutron stars remain, but the predicted rates of gravitational wave detections for the Advanced LIGO detector tend to be or order 1-100 yr$^{-1}$ \cite{Abadie2010,Faber2012}.

Gamma-ray bursts (GRBs) are extremely energetic events (releasing $\gtrsim 10^{50}$ergs worth of $\sim$ MeV $\gamma$-rays) at cosmological distances; short GRBs (sGRBs), lasting typically $\lesssim 2$s, have been are strong candidates to be the EM signatures of binary neutron star mergers\cite{Eichler1989} (for a review of evidence linking sGRBs to NS-NS mergers, see \cite{Berger2013}). 

%====================================== FIGURE 10 ===============================

\begin{figure}
\resizebox{0.5\textwidth}{!}{
  \includegraphics{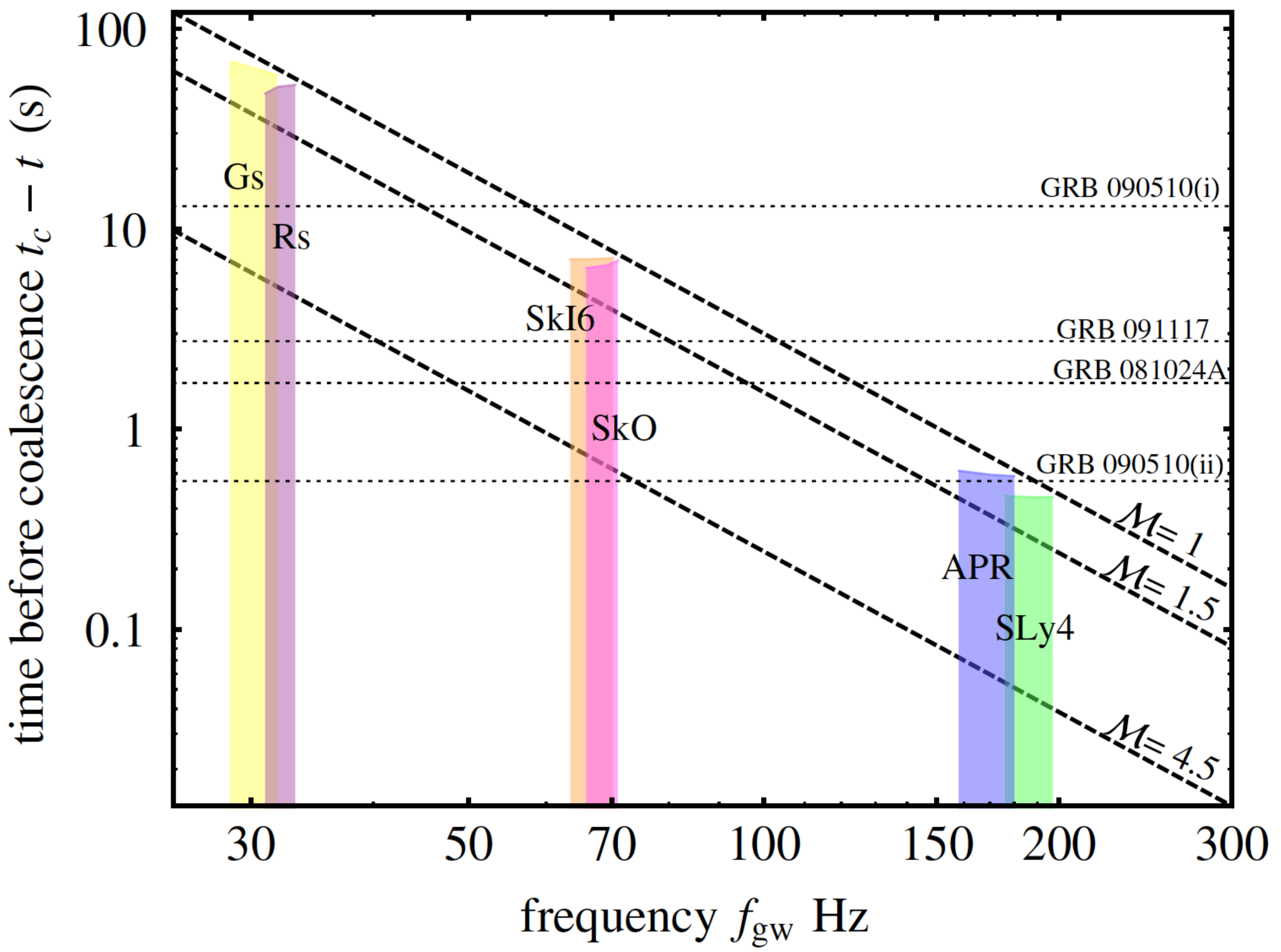}}
\caption{Pre-merger time $t_{\rm c}$ versus frequency of gravitational waves emitted by the binary system $f_{\rm GW}$. Binaries follow trajectories indicated by the bold dashed lines for three different chirp masses $\mathcal{M}$. The observed pre-sGRB times for the EM flares are shown by the horizontal dotted lines. The predicted times for the crust shattering are shown by the colored bars for 6 different EOSs. The EOSs have the following values for the slope of the symmetry energy $L$. Gs: $L$=93.31 MeV, Rs: $L$=86.39 MeV, SkI6: $L$=59.24 MeV, SkO: $L$ = 79.14 MeV, APR: $L$=59.63 MeV, SLy4: $L$ = 45.94 MeV \cite{Dutra2012}. Taken from \cite{Tsang2012}.}
\label{fig:10}  
\end{figure}

%==============================================================================

Five sGRBs have indications of precursor gamma-ray flares flares of order 1 - 10s prior to the main burst \cite{Troja2010}. The mechanism for such precursor events is currently open to interpretation, but one suggestion is that the emission is associated with the catastrophic mechanical failure of the neutron star crusts induced by dynamical tidal interactions between the neutron stars in the immediate run-up to merger.

Crust failure occurs when the breaking strain $\epsilon_{\rm b} \approx 0.1$ is exceeded \cite{Chugunov2010}, which can occur when the tidal deformation reaches a certain value. The failure of the crust induced purely by the dynamical tidal deformation of the star is unlikely to cause sudden catastrophic failure, with different regions of the crust exceeding the breaking strain at different times \cite{Penner2012}. If the released mechanical energy is deposited as heat in the crust, it will not diffuse to the surface in time to leave an observable EM signature prior to merger; the most energetic failure might excite seismic crustal modes which couple to the magnetosphere, leading to another avenue for EM emission, but in this scenario this would occur less than 1s before crustal failure.

In the model of \cite{Tsang2012}, the dynamical tidal interaction between one neutron star in its companion resonates a particular global oscillation mode, and resonantly shatters the crust. The candidate mode with a natural frequency commensurate with the orbital frequency of the binary system between 1 and 10 seconds before merger, $\sim 100$Hz, is a quadrupolar crust-core interfacial mode, and has a strong terrestrial analogue in a Scholte wave - a wave which travels along an elastic solid - liquid interface. The natural frequency of such a mode depends strongly on the symmetry energy, both through the dependency on the stellar radius and on the crust-core transition density $n_{\rm cc}$ (and the crustal shear modulus there $\mu_{\rm cc}$). 

%\textbf{$\mu/P$ at base of spherical-pasta layer}. A reduction in the shear modulus by a factor of three would reduce the interfacial frequency by $\sqrt{3}$, moving the bars down and the slope of the symmetry energy range down.

\begin{table*}
\begin{center}
\caption{Ranges of the slope of the symmetry energy $L$(MeV) inferred from modeling neutron star observables  \label{Tab1}}
\begin{tabular}{lclc}
\hline
Observable & $L$ (MeV) & Specific (general) conditions/caveats & Reference \\

\hline
Cooling rate of Cas A  & $\lesssim$ 70 & No pasta cooling processes & \cite{Newton2013} \\
neutron star & $\lesssim$ 45 & Pasta cooling processes active and unsuppressed by crust superfluity & \\
%&& & \\
&& (Minimal cooling paradigm; range of $L$ contingent on atmosphere model)  & \\
\hline
Limiting spin period of high & $\lesssim$ 80 & Magnetic field decay from highly resistive pasta layer, not& \cite{Pons2013} \\
magnetic field X-ray pulsars && high resistivity of an amorphous/heterogeneous inner crust& \\
\hline
Vela pulsar glitches & $\gtrsim$ 100 & Full crustal entrainment, very weak crust-core coupling. & \cite{Hooker2013} \\
&&  Glitch mechanism might involve angular momentum transfer  &\\
&& from core components. &\\
\hline
QPOs in X-ray tails of & $\lesssim$ 60 & Calculated frequencies fall in range of potential observed fundamental & \cite{Gearheart2011} \\
giant flares from SGRs && frequencies; consistent crust-core EOS;  &\\
&& limiting superfluid, pasta effects included &\\
& $\gtrsim$ 50 & Exact matching of fundamental mode with lowest observed frequency  & \cite{Sotani2012} \\
&& QPO; inconsistent crust, core models; no superfluid effects; &\\
& 100 $\lesssim L \lesssim$ 130 & Exact matching of all observed frequency with crust modes; & \cite{Sotani2013a} \\
&& inconsistent crust, core models; superfluid effects included & \\
& 58 $\lesssim L \lesssim$ 85 & As above, but with the 2nd lowest observed frequency from SGR1806-20  & \cite{Sotani2013b} \\
&&  omitted in mode indentification &\\
&& (Alfven wave coupling to crust modes ignored. & \\
&&  Low frequency modes could be explained by pure Alfven modes.) &\\
\hline
Limiting spin-up & $\lesssim$ 65& Consistent crust-core EOS; viscous & \cite{Wen2012} \\
frequency of  & & dissipation at crust-core boundary & \\
millisecond pulsars & $\gtrsim 50$ &Inconsistent crust-core model; viscous & \cite{Vidana2012} \\
&& dissipation throughout entire core & \\
&& (Crust not perfectly rigid. r-mode saturation might allow stars to spin&\\
&&  -up into instability window. Superfluid, exotic shear viscosity sources  &\\
&&   ignored. Alternative physical mechanisms that limit spin-up are possible.) &\\
\hline
Observed occurrence & 60 $\lesssim L \lesssim$ 80 & Inconsistent crust-core EOS. Observational interpretation & \cite{Tsang2012} \\
times of  precursor  && of pre-cursor gamma ray signals tentative. &\\
$\gamma$-ray flares before sGRBs &&& \\
\hline
 
\end{tabular}

\end{center}
\end{table*}

Fig.~10 shows the pre-merger time versus the frequency of gravitational waves emitted by the time-varying system mass quadrupole, $f_{\rm gw}$, which is twice the orbital frequency \cite{Tsang2012}. The light-dashed lines show the pre-sGRB time for 4 precursor flares. The bold dashed lines show the evolution in this parameter space of binary systems with a representative range of chirp masses $\mathcal{M} = M_1^{3/5} M_2^{3/5} / (M_1 + M_2)^{1/5}$. The horizontal bars represent results of calculations of the interfacial mode frequencies for 6 different EOSs, and are positioned at the frequency of the interface mode, and their heights show the time before merger at which the frequency of their dynamical tides (the orbital frequency ) will come into resonance with the interface modes and shatter the crust. The values of $L$, given in Fig.~10, span the range $\approx 45 - 90$ MeV, sufficient variation to change to pre-merge time by nearly two orders of magnitude. The pre-merger times roughly correlate with $L$, with higher $L$ values tending to give larger pre-merger times. It is important to note that the crust-core transition density was fixed at $n_{\rm cc} = 0.065$fm$^{-3}$ in this work, rather than allowing it to additionally depend on the EOS. Therefore the $L$ dependence emerges from its correlation with radius $R$. Larger $L$ gives a large star and thicker crust, leading to lower interfacial crustal frequencies which will come into resonance with the tidal forces at earlier times.

Even accepting the scenario, there is significant uncertainty, but in this analysis one can claim that the best agreement with the observed precursor flare times comes in the region $f_{\rm gw} \approx  60-180$Hz, a region which includes the SkI6, SkO and APR EOSs. These EOSs span a range of $L \approx 60 - 80 MeV$. \\

\noindent \textbf{\emph{Constraint:}} $60 \lesssim L\lesssim 80$ MeV.

\noindent \textbf{\emph{Caveats:}} Crust not calculated consistently with core EOS. The proposed mechanism for electromagnetic radiation as a result of catastrophic crustal failure needs a more quantitative grounding. The observational interpretations of EM precursors is still quite tentative.

\section{Conclusions}

We have reviewed recent work modeling neutron star phenomena in which crust physics plays a significant role with a view to exploring the sensitivity of the phenomena to the density slope of the symmetry energy $L$. These various models have been confronted with six sets of observations, taken at various stages of the neutron star's life from $\approx$ 300 yrs after its birth to a few seconds before its death, $\sim10^9$yrs later. Table~1 summarizes all the constraints on $L$ obtained from these analyses. 

We start by pointing out that these diverse phenomena present a variety of challenges, both in the interpretation and in the modeling required to understand them. There are many points in any one of the analyses presented in which significant source of uncertainties can enter. Many of the observations are subject to possible systematic error. The theoretical models presented are far from complete, often neglecting important physics such as magnetic fields in order that they be tractable; many of them should be viewed as not much more than toy models. Finally, the derivation of constraints on neutron star physics from the application of theoretical models to observations is of course contingent on the correct interpretation of the observations, something that is still open to debate in many cases.

The sensitivity to the symmetry energy in all these cases emerges mainly through the sensitivity of the crustal volume and mass, and that of the nuclear pasta phases, to the density expansion parameters of the symmetry energy, particularly the first-order parameter $L$. There is also some contribution from the dependence of the crustal composition on the symmetry energy. Various other important microscopic quantities such as the pasta shear modulus, breaking strain, crustal neutron superfluid gap and entrainment, thermal and electrical transport properties have been taken to be fixed. Their dependence on the symmetry energy needs to be quantified if we are to place rigorous astrophysical constraints on nuclear interactions.

Nevertheless, no matter how simplified the models, their demonstrated sensitivity to the symmetry is unlikely to be significantly reduced upon the inclusion of more realistic physics. What these studies demonstrate is that there exist a wealth of observations of diverse phenomena that can help constrain the symmetry energy in the future. Furthermore, as constraints on the symmetry energy become narrower, such modeling can offer consistency checks and test the validity of a particular model or observational interpretation. For example, if a strict constraint of $L<$70 MeV is obtained by some future experiment, this places serious doubts on the current pulsar glitch paradigm, and indicates that the model of crustal modes or the identification of crustal modes in \cite{Sotani2013b} is incomplete. Thus, as well as astrophysical studies placing constraints on nuclear physics, in the future we can expect nuclear physics constraints to place ever tighter constraints on the theoretical models and interpretation of astrophysical phenomena. We can look forward to an ever tighter interplay between nuclear physics and astrophysics.

\section{Acknowledgments}
This work is supported in part by the National Aeronautics and Space Administration under grant NNX11AC41G issued through the Science Mission Directorate, 
the National Science Foundation under Grants No. PHY-1068022, 
the REU program under grant no. PHY-1062613, 
CUSTIPEN (China-U.S. Theory Institute for Physics with Exotic Nuclei) under DOE grant number DE-FG02-13ER42025, 
the National Natural Science Foundation of China under Grant No.s 10947023,11275073 and 11320101004, 
and Fundamental Research Funds for the Central Universities in China under Grant No. 2014ZG0036.


%\begin{figure*}
%\vspace*{5cm}       % Give the correct figure height in cm
%\end{figure*}
%\begin{table}
%\begin{tabular}{lll}
%\hline\noalign{\smallskip}
%\end{tabular}
% Or use
%\vspace*{5cm}  % with the correct table height
%\end{table}
%
% BibTeX users please use
% \bibliographystyle{}
% \bibliography{}

\end{document}